\documentclass[prd,aps,
showpacs,
nofootinbib,floatfix,superscriptaddress]{revtex4}
\usepackage{graphicx}
\usepackage{epsfig,epic}
\usepackage{rotating}
\usepackage{amssymb}
\usepackage{dsfont}
\usepackage{psfrag}
\usepackage{amsmath}

\newcommand{\beq}{\begin{equation}}
\newcommand{\eeq}{\end{equation}}
\newcommand{\beqs}{\begin{eqnarray}}
\newcommand{\eeqs}{\end{eqnarray}}

\def\hbar{\hspace{0pt}\raisebox{1pt}{$-$} \hspace{-7pt} h}

\def\di{\mbox{d}}
\def\r{\rho}

\newcommand{\be}{\begin{equation}}
\newcommand{\ee}{\end{equation}}
\newcommand{\bea}{\begin{eqnarray}}
\newcommand{\eea}{\end{eqnarray}}

\def\lbldef#1#2{\expandafter\gdef\csname #1\endcsname {#2}}

\def\href#1#2{#2}

\newcommand{\ber}{\begin{eqnarray}}
\newcommand{\eer}{\end{eqnarray}}

\newcommand{\beqar}{\begin{eqnarray}}

\newcommand{\eeqar}{\end{eqnarray}}

\newcommand{\dsl}
  {\kern.06em\hbox{\raise.15ex\hbox{$/$}\kern-.56em\hbox{$\partial$}}}

\newcommand{\eeqarr}{\end{eqnarray}}
\newcommand{\ZZ}{{\rm \kern 0.275em Z \kern -0.92em Z}\;}

\def\CC{{\mathchoice
{\rm C\mkern-8mu\vrule height1.45ex depth-.05ex
width.05em\mkern9mu\kern-.05em}
{\rm C\mkern-8mu\vrule height1.45ex depth-.05ex
width.05em\mkern9mu\kern-.05em}
{\rm C\mkern-8mu\vrule height1ex depth-.07ex
width.035em\mkern9mu\kern-.035em}
{\rm C\mkern-8mu\vrule height.65ex depth-.1ex
width.025em\mkern8mu\kern-.025em}}}

\def\RR{{\rm I\kern-1.6pt {\rm R}}}

\def\ZZ{{\rm Z}\kern-3.8pt {\rm Z} \kern2pt}
\def\IB{\relax{\rm I\kern-.18em B}}
\def\ID{\relax{\rm I\kern-.18em D}}
\def\II{\relax{\rm I\kern-.18em I}}
\def\IP{\relax{\rm I\kern-.18em P}}

\newcommand{\bear}{\begin{eqnarray}}
\newcommand{\eear}{\end{eqnarray}}

\def\to{\rightarrow}

\def\to{\rightarrow}

\def\a{\alpha}

  \def\w{\omega}
\def\r{\rho}                                     
\def\s{\sigma}                                   
\def\t{\tau}

\def\lab{\label}
\def\6{\partial}

\def\bea{\begin{eqnarray}}
\def\eea{\end{eqnarray}}

\def\beqx{\begin{displaymath}}
\def\eeqx{\end{displaymath}}

\newcommand{\bmat}{\left(\begin{array}}
\newcommand{\emat}{\end{array}\right)}

\def\a{\alpha}

\def\r{\rho}
\def\s{\sigma}
\def\t{\tau}

\def\bo{{\raise-.3ex\hbox{\large$\Box$}}}               
\def\face{{\raise.2ex\hbox{$\displaystyle \bigodot$}\mskip-2.2mu \llap {$\ddot
        \smile$}}}                                   
\def\>{\rangle}                                      
\def\<{\langle}                                      

\def\leftrightarrowfill{$\mathsurround=0pt \mathord\leftarrow \mkern-6mu
        \cleaders\hbox{$\mkern-2mu \mathord- \mkern-2mu$}\hfill
        \mkern-6mu \mathord\rightarrow$}        
\def\dvec#1{\vbox{\ialign{##\crcr
        \leftrightarrowfill\crcr\noalign{\kern-1pt\nointerlineskip}
        $\hfil\displaystyle{#1}\hfil$\crcr}}}           

\def\-{\hphantom{-}}

\begin{document}
\title{Wilson Loops in string duals of Walking and Flavored Systems.}

\author{Carlos N\'u\~nez, Maurizio Piai and Antonio Rago}
\affiliation{Swansea University, School of Physical Sciences,
Singleton Park, Swansea, Wales, UK}

\date{\today}

\begin{abstract}
We consider the VEV of Wilson loop operators by studying the behavior of 
string probes  in solutions of Type 
IIB string theory generated by $N_c$ $D5$ branes wrapped on an $S^2$ internal manifold.
In particular, we focus on solutions to the background equations
 that are dual to field theories with a 
{\it walking gauge coupling} as well as for flavored systems.
We present in detail our 
walking solution and emphasize various general aspects of the procedure to 
study 
Wilson loops using string duals. We discuss the special features  that 
the strings show 
when probing the region associated with the walking of the field 
theory coupling. 
 \end{abstract}


\maketitle

\tableofcontents
\newpage

\section{Introduction}
In this paper we want to study the behavior of non-local operators 
of gauge theories, making use of the gauge-string correspondence.
We are in particular interested in a specific class of supergravity solutions that are closely 
related to what goes under the name of {\it walking} in the field theory 
language.
In the introduction we summarize the basic notions and ideas 
that will feature prominently in the paper:  the treatment of Wilson loops in the gauge-string
correspondence, the concepts of confinement and screening 
in gauge theories and the meaning of walking dynamics (with a view on 
its role within dynamical electro-weak symmetry breaking).

It must be stressed that we do not know the precise nature of the field-theory dual 
of some of the examples we are going to consider in the body of the paper.
The Wilson loop studied here is a very important quantity, that may help 
identifying such dual theory.

\subsection{AdS/CFT and Wilson loops}

According to general ideas of holography and more concretely to the 
Maldacena conjecture \cite{Maldacena:1997re}, a quantum conformal field 
theory in dimension $d$ is equivalent to a quantum theory in $AdS_{d+1}$ 
space. 
In general, the idea is that local operators in the CFT
couple to fields in the $AdS$ side, in such a way that
correlators of conformal fields are related to amplitudes in the 
quantum theory in $AdS$, as explained in \cite{Gubser:1998bc}. 
For instance, since the 
CFT-side of the equivalence must contain the energy-momentum tensor 
$T_{\mu\nu}$
among its operators, there must be a field on the $AdS$-side that 
couples to it. Since this field is the graviton, then the theory on 
$AdS_{d+1}$ space must be a quantum theory of gravity.

One can use the correspondence to study non-local operators
on the CFT-side. In particular, if the 
field theory is a pure Yang-Mills theory, 
an  example of such operator is the Wilson loop~\cite{Wilson:1974sk}. 
These objects couple to extended objects, excited in the $AdS$ side of the correspondence.
The Wilson loops (path ordered exponentials of the holonomy of the gauge 
field along a curve ${\cal C}$) are one of the most interesting 
observables of such a gauge theory,
\beq
W({\cal C})\equiv \frac{1}{N_c} Tr P e^{i\oint_{\cal C}A_\mu dx^\mu}.
\label{wilson1}
\eeq
The loop itself and products of them provide a basis of gluonic gauge 
invariant 
operators.

The Wilson loop along a  curve ${\cal C}$ is computed in the dual string 
theory by calculating the action of a string bounded by  ${\cal C}$ at 
the boundary of the $AdS$ space. More concretely\footnote{The fundamental string is actually dual to the generalized Wilson loop,  containing a term that couples the coordinates of the internal space with the scalars of the brane. This is due to the fact that the string ending on the brane is source of the electric field, generating $A_\mu$ but also of the scalars on the brane, from which is pulling. See \cite{Maldacena:1998im} for a clear discussion of this.},
\beq
\left\langle \frac{}{}W({\cal C})\right\rangle= \int_{\partial F({\cal C})} {\cal D} F e^{- S[F]}
\label{definitionwilson}
\eeq
where $F$ denotes all the fields of the string theory and $\partial F$ their 
boundary values. A good approximation to this path integral is by steepest 
descent.  The Wilson loop is then related to the area of the minimal 
surface  bounded by 
the curve ${\cal C}$, 
spanned by classical string  configurations
(with Nambu-Goto action $S_{NG}$) that 
explore the bulk of $AdS$.

All of  this was first proposed ten years ago in 
\cite{Maldacena:1998im}.
In the meantime, this proposal motivated lots of developments, see 
\cite{bunchwilson} for beautiful and influential papers on this line.
See also~\cite{Sonnenschein:1999if}
for a review.
In particular, the ideas and techniques of the original $AdS/CFT$ correspondence have been
assumed to generalize to a large class of systems,
 and have been used to relate field theories that are not conformal
 (and hence more closely related to phenomenological applications)
  with backgrounds that are not $AdS$, extending the framework to what is more 
appropriately referred to as {\it gauge-string} duality.

\subsection{Confinement and Screening}

In its original definition~\cite{Wilson:1974sk}, the Wilson loop
computes the phase factor associated to a closed  trajectory for a very 
massive quark in the fundamental representation (it  can also be generalized
to other representations).
The quark-antiquark (static) potential can be read from the VEV of the Wilson loop.
Choosing a rectangular loop of 
sides $L_{QQ}, T$, in the first approximation for large times $T\to\infty$,
\beq
\left\langle \frac{}{}W({\cal C})\right\rangle\sim e^{- E_{QQ} T} \,,
\eeq
where $E_{QQ}$ is the quark-pair energy. In the limit of 
large 't Hooft coupling,  the 
steepest descent approximation mentioned above yields the identification
\beq
\left\langle \frac{}{}W({\cal C})\right\rangle\sim e^{- E_{QQ} T}\sim e^{-S_{NG}}\,.
\eeq

The description of the Wilson loop in QCD in terms of  a string partition 
function is not new. Well before its modern formulation, 
the ideas behind  gauge/string correspondence have been used to
show that the potential of a  quark-antiquark pair
separated by a distance $L_{QQ}$ gets a correction of the form 
$\frac{c}{L_{QQ}}$ due to quantum fluctuations of the Nambu-Goto action~\cite{Luscher:1980fr}.

The definition of confinement
we will adopt is the following. Consider a $SU(N_c)$ gauge theory with 
matter fields in generic representations. We decouple (make infinitely massive) 
all fields with non-zero N-ality,  and then introduce a single particle-antiparticle pair
of non-dynamical fields with non-zero N-ality as a test probe for the system
(effectively, the pair dynamics is quenched).
We then  compute the work needed to separate 
the particle-antiparticle test pair up to a distance $L_{QQ}$. 
If the work approaches $E_{QQ}\simeq \sigma L_{QQ}$ for large separations, the theory is 
{\it confining}\footnote{Another equivalent way of defining confinement is by 
computing the VEV of the Polyakov loop, that if vanishing indicates a 
confining theory. Also, the 
perimeter law of a 't Hooft loop indicates 
confinement.}. The quantity  $\sigma$ is a representation dependent constant, the 
string tension. 

\subsection{Walking technicolor}

Walking technicolor~\cite{WTC} is a framework within which
the phenomenological difficulties of dynamical electro-weak symmetry 
breaking might find a very natural and elegant dynamical solution,
thanks to the fact that, in contrast to QCD-like theories,
 the guidelines provided by  naive dimensional analysis are violated.
 This is because of large anomalous dimensions controlling the dynamics 
 over a large regime of energies.
The word {\it walking} refers to the fact that  the new dynamics is strongly coupled 
over a large range of energies, where its fundamental coupling  exhibits 
a  $\beta$-function which is anomalously small in respect to the coupling itself.
A behavior of this type is
 expected in theories which flow onto strongly-coupled IR fixed-points,
and it is reasonable to assume that it persists
 also when such IR fixed points are only approximate, though 
 in this case a degree of ambiguity as to the meaning of {\it approximate} invites some caution.

Besides being affected by 
the usual calculability limitations due to the strong coupling (as in QCD and in QCD-like technicolor),
the walking dynamics itself makes this framework very hard to work with.
New, non-perturbative instrument are needed in order to understand 
the (effective) field theory properties of a
walking theory.
Very recent years saw a lot of progress towards a better understanding of walking dynamics
both from the lattice~\cite{lattice} and thanks to ideas borrowed from  the gauge-string duality.
See for instance~\cite{AdSTC} for a list of references focused 
on the precision electro-weak  parameters.

In this paper, we will use the word {\it walking} to refer to backgrounds in which 
there exists a interval in the $\r$ radial direction
over which the geometry is determined by some coupling that shows 
an evolution with the radial coordinate that is anomalously slow. 
This agrees with  the standard definition of walking, 
where the beta-function of the fundamental gauge-coupling is, over some range of
energy, anomalously small in comparison with what expected from the strength of the actual coupling itself.
Not all couplings show this behavior: this also agrees with standard definitions,
in the sense that relevant operators must be present in order to 
drive the RG flow away from a possible IR fixed point, and scale invariance is 
present only in the sense that in the walking region the relevant couplings are 
so small that their effect can be neglected.
However, our definition is significantly less restrictive that the usual one: we do not require the presence of an actual fixed point
in the flow, and correspondingly we do not have an approximately AdS background, nor can we recover 
an AdS background by dialing some parameter to some special value.

\subsection{General idea:  string-theory  as a laboratory for walking dynamics.}
Motivated by the difficulties described in the previous subsection, in ref.~\cite{Nunez:2008wi} a  more 
general program is proposed,
based upon gauge/string correspondence in order to
go beyond the low-energy effective field theory description.
The proposal is to study the  dynamics of theories that yield
walking behavior, but that 
are not necessarily related to  the electro-weak symmetry.
In short, one would like to use string theory 
as a laboratory in which to study the general properties of walking dynamics by itself, 
in isolation from 
its complicated realization within an explicit 
model of dynamical electro-weak symmetry breaking.
In ref.~\cite{Nunez:2008wi}, it is shown that 
in the context of Type IIB string theory on a background generated by a stack of
$N_c$ $D5$ branes,
there exists a very large class of solutions to the background equations
for which a suitably defined gauge coupling exhibits the basic properties of 
a putative walking theory. The running of the gauge coupling 
flattens over a large range of 
intermediate energies, but restarts at low energies, until the space 
ends into a (good) singularity in the deep IR, 
so that no exact IR fixed point exists.
The fact that  this is not a walking technicolor theory (there is no
electro-weak symmetry in the set-up, and hence no mass generation in the usual
sense), together with the large-$N_c$ expansion, 
yields the advantage that we avoid the
complications due to mixing of weakly-coupled and strongly-coupled 
properties of the theory.
 For instance, 
the spectrum of the spin-0 sector of the theory
can be studied, and has been studied~\cite{Elander:2009pk}, yielding remarkable surprises.

In this paper we take a further step in this direction, by studying the 
behavior of the Wilson loop in backgrounds of this class.
As we will explain, we can use the techniques developed 
in the context of gauge-string duality, by studying the 
background with 
a probe string. In particular, we will study Wilson loops in the dual 
{\it walking} QFT.
For technical reasons that will be explained in the body of the paper,
in order for this program to be carried out we will also need to generalize further the class of
backgrounds in~\cite{Nunez:2008wi}. These new solutions have been 
 already introduced in~\cite{Elander:2009pk}. 
We explain here in deeper detail how to generate them, characterize them,
 and relate them to the literature.

\subsection{Outline}

The paper is organized as follows: we set up notation and introduce a set 
of important ideas in section \ref{general} revising the bibliography and 
adding important new ingredients and derivations. Then we apply these to 
well-established examples of gauge-string duality in section 
\ref{examplesunderstood}, providing a simple 
and compact set of exercises that are intended to yield some guidance 
in the following sections, in which the dynamics is far from well-understood. 
Section \ref{walkingsolutionsxx} presents our 
new {\it walking} solution and a study of the dynamics
of the Wilson loop as a function of the length of the walking region. 
Section \ref{flavoredbackgroundsxx} studies the results derived in section 
\ref{general} when applied to background that encode the dynamics of 
fundamental fields. We then conclude in section \ref{conclusionszz}.

\section{General theory}\label{general}
In this section we present general results for Wilson loops,  computed 
using the ideas of \cite{Maldacena:1998im}. Some of the results here have 
been derived long ago (see for example \cite{Kinar:1998vq}), but our 
approach will be 
different and some new and useful points will be specially emphasized.

We study the action for a string in a background of the generic form
\beq
ds^2=-g_{tt} dt^2+ g_{xx} d\vec{x}^2+ g_{\r\r}d\r^2+ g_{ij}d\theta^i 
d\theta^j .
\label{back}
\eeq
We assume that the functions $(g_{tt}, g_{xx}, g_{\r\r})$ depend only on the 
radial coordinate $\r$. 
By contrast, $g_{ij}$ for the internal (typically compact) space can 
also depend on other coordinates. 
Whatever are the internal coordinates, they will play no role in what 
follows.
This is because we will choose a configuration for a probe string that is 
not excited on the $\theta^i$ directions, hence in what follows,  we will 
ignore  the 
internal space \footnote{Strings or other 
objects that extend in the internal space filling part of it can be treated as 
an effective string, analogous to the one we are studying. If these objects are allowed to vibrate in 
the internal space, then a generalization of the present treatment should  be 
done.}. 
\subsection{Equations of motion}\label{eqsofmotion}
The configuration we  choose is,
\beq
t=\tau,\;\;\;\; x=x(\sigma),\;\;\;\; \r=\r(\sigma).
\label{ansa}
\eeq
and compute the Nambu-Goto action
\beq
S= \frac{1}{2\pi \alpha'}\int_{[0, T]} d\tau \int_{[0,2\pi]} d\sigma 
\sqrt{-\det G_{\alpha\beta}}. 
\label{ng}
\eeq
The induced metric on the 2-d world-volume is 
$G_{\alpha\beta}=g_{\mu\nu}\partial_\alpha X^\mu \partial_\beta X^\nu$, 
where
\beq
G_{\tau\tau}=-g_{tt},\;\;\; G_{\sigma\sigma}=g_{xx}(\frac{dx}{d\sigma})^2+ 
g_{\r\r} 
(\frac{d\r}{d\sigma})^2 \,.
\label{zxzx}
\eeq
Defining for convenience
$ f(\r)^2\equiv g_{tt}g_{xx},\; g(\r)^2= g_{tt}g_{\r\r}$, 
the Nambu-Goto action  is
\beq
S= \frac{T}{2\pi \alpha'} \int_{0}^{2\pi} d\sigma
\sqrt{f^2 x'(\sigma)^2 + g^2 \r'(\sigma)^2}
\,\equiv\,
 \frac{T}{2\pi \alpha'} \int_{0}^{2\pi} d\sigma L\,.
\label{ng2}
\eeq
Notice that we consider the situation in which the string does not 
couple to a NS $B$-field. 

We first compute the Euler-Lagrange equations  from Eq.~(\ref{ng}) and  
then we specify them  for the ansatz in Eq.~(\ref{ansa}). We get that for the 
$(t, x, \r)$ coordinates the eqs. of motion read, respectively,
\bea
& & \partial_\tau \Big[\frac{1}{L} (f^2 x'^2 + g^2 \r'^2)  
\Big]=0\,,\nonumber\\
& & \partial_\sigma \Big[\frac{1}{L} f^2 x'   \Big]=0\,,\label{eqtxraa}\\
& & \partial_\sigma \Big[\frac{1}{L} g^2 \r'   \Big]=\frac{1}{L}(x'^2 f f' 
+ \r'^2 g g')\nonumber\,,
\eea
where  $X^{\prime}=\frac{d X(\s)}{d\s}$ for any function $X$.

\begin{figure}[htpb]
\begin{center}
\includegraphics[height=7cm]{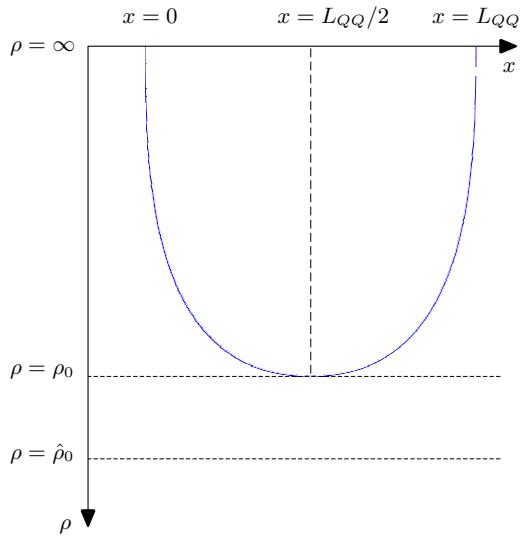}
\caption{Setting of the string.}
\label{Fig:stringset}
\end{center}
\end{figure}

The first equation in (\ref{eqtxraa}) is solved because we assume a 
background metric independent of time (we consider the system at the equilibrium). 
The second equation in 
(\ref{eqtxraa}) is solved if the quantity inside brackets is a 
constant (that 
we call $C$), which implies that,
\beq
\label{drdx2}
\frac{d\r}{d\sigma}=\pm\Big(\frac{dx}{d\sigma}\Big)\Big(\frac{f(\r)}{C 
g(\r)}\Big)\sqrt{f^2(\r)- C^2}
\eeq
One can check that the third equation in (\ref{eqtxraa}) is solved if 
Eq.~(\ref{drdx2}) is imposed. Then, we need to work with just one equation.  
Defining 
\beqs
V_{eff}(\r)&\equiv&\frac{f(\r)}{C g(\r)}\sqrt{f^2(\r)- C^2}\,,
\label{nnzz}
\eeqs
we write it as
\beq
\label{drdxfinal}
\frac{d\r}{d\sigma}=\pm\frac{dx}{d\sigma} V_{eff}(\r)\leftrightarrow \frac{d\r}{d x}= 
\pm V_{eff}(\r)\,.
\eeq
Another way to arrive to the last version of Eq.~(\ref{drdxfinal}) is to 
consider a restricted 
ansatz for the string configuration $
[t=\tau,\; x=\sigma,\;\; \r=\r(\sigma)]$ 
and use the conserved Hamiltonian derived from Eq.~(\ref{ng2}) to get an 
expression for $\r(\sigma)$ that is precisely Eq.~(\ref{drdxfinal}).\\
The kind of solution we are interested in can be depicted as follows: a string that hangs from infinite radial position at $x=0$ and drops down towards smaller $\r$ as $x$ increases. Once it arrives at the smallest $\r$ compatible with the solution, namely $\r_0$, it starts growing in the radial direction up to infinite $\r$ where $x=L_{QQ}$, see Fig.~(\ref{Fig:stringset})\footnote{For future reference we refer to the lowest end of the radial coordinate as $\hat{\r}_0$.}. 
This means that in the two distinct regions $x<L_{QQ}/2$ and $x>L_{QQ}/2$ the equations of motion will differ only in a sign
\begin{eqnarray}
\label{Eq:eqofmot}
x<\frac{L_{QQ}}{2} && \frac{d\r}{d x}= -V_{eff}(\r)\nonumber\\
x>\frac{L_{QQ}}{2} && \frac{d\r}{d x}= V_{eff}(\r)
\end{eqnarray}
We can now formally integrate the equations of motion 
\beqs
x(\r)=
\begin{cases}
\int_{\r}^{\infty}\frac{d\r}{V_{eff}(r)}&\quad x<\frac{L_{QQ}}{2}\\
L_{QQ}-\int_{\r}^{\infty}\frac{d\r}{V_{eff}(r)}&\quad x>\frac{L_{QQ}}{2}
\end{cases}
\eeqs
or more compactly
\beq
\label{Eq:eqofmotcompact}
\left|x(\r)-\frac{L_{QQ}}{2}\right|=\int_{\r_0}^{\r}\frac{dr}{V_{eff}(r)}\,.
\eeq
In what follows we will use only one of the solutions in Eq.~(\ref{Eq:eqofmot}) unless explicitly noted.

\subsubsection{Boundary conditions}\label{boundaryconds}
We need to specify the boundary conditions for  the string in 
Eq.~(\ref{ansa}). This is an open string, vibrating in the bulk of a closed 
string background. Following the ideas of 
\cite{Maldacena:1998im}, we add a D-brane at a very large radial distance 
where the open string will end. This string will then satisfy a Dirichlet 
boundary condition at $\r\to\infty$. This means that for large values of 
the radial coordinate
$\frac{dx}{d\sigma}$ must vanish. 
The only way of satisfying the equation of motion in Eq.~(\ref{drdx2}) for $\rho\to\infty$, given that the left hand side has to be non vanishing, is to have a divergent $V_{eff}(\r)$:
\beqs
\label{Eq:UVboundary}
\lim_{\r\rightarrow \infty} V_{eff}(\r)&=&\infty\,.
\eeqs
 This implies that there are restrictions on the asymptotic behavior 
 of the background functions [$f(\r), g(\r)$] in order for the string proposed in 
Eq.~(\ref{ansa}) to exist. We will come back to this in the following 
sections.
Before proceeding, a brief digression is needed.
When studying Eqs.~(\ref{drdxfinal})-(\ref{Eq:UVboundary}) the reader may 
find unsatisfactory that the restriction we have proposed above, namely
\beq
\left.\frac{d\r}{dx}\right|_{\r\to\infty}=\left.V_{eff}\right|_{\r \to \infty}\to \infty
\eeq
looks  dependent of our choice of the radial coordinate.
This is actually not the case, because we could rewrite this restriction in a 
more covariant form in the following way.
We define a couple of vectors~\footnote{We thank Johannes Schmude for the 
discussions that lead to this.}  in the 
$x,\r$
directions,
\beq
\vec{v}^x\equiv\frac{dx}{d\sigma}\partial_x,\;\;\; 
\vec{v}^\r\equiv\frac{d\r}{d\sigma}\partial_\r\,,
\eeq
compute the ratio $\mu$ of their norms and impose that this is 
divergent on the surface ${\cal D}$ on which
the string has to satisfy the Dirichlet condition:
\beq
\mu=\left. \frac{g_{\r\r} (\frac{d\r}{d\sigma})^2}{g_{xx} 
(\frac{dx}{d\sigma})^2}\right|_{{\cal D}}\to\infty.
\eeq
Combining this with Eq.~(\ref{drdxfinal}), and evaluating at the boundary
\beq
\mu=\left.\frac{g_{\r\r}}{g_{xx}}V_{eff}^2\right|_{{\cal D}}=\frac{f^2({\cal D})-C^2}{C^2}
\to\infty .
\eeq
This last 
expression is free of coordinate ambiguities, as it comes from operating 
with invariants (norms of vectors). It is however easier to 
work within a specific choice of coordinates, 
which we will do in the body of the paper.
As we will see explicitly, 
this occurs in the examples we will study below. 

\subsubsection{Turning points}
Once Eq.~(\ref{Eq:UVboundary}) is satisfied the string will move to smaller values of the radial coordinate down to a turning point $\r_0$ where the quantity $\frac{d\r}{dx}(\r_0)=0$, i.~e. $V_{eff}=0$.
In principle, there could be points where $V_{eff}$ vanishes because of either isolated zeros of $f(\r)$, or diverging point of $g(\r)$. However, we will not consider this kind of inversion points, since we are interested in solutions of the equations of motion that allow the string to probe the entire radial direction. Hence the turning point can be placed in any possible $\r_0$, with $\hat{\r}_0 < \r_0 < \infty$ (where $\hat{\r}_0$ is the end of the space).
Thus we will restrict ourselves to forms of $V_{eff}$ where the inversion point is given by imposing $C=f(\r_0)$. Furthermore, in the next section we will also impose that the envelop of the string is convex in a neighbourhood of $\r_0$, hence insuring that gauge theory quantities like the 
separation between the pair of quarks and its energy will be continuous 
functions of $\r_0$.
It is then clear from Eq.~(\ref{drdxfinal}) 
that $V_{eff}(\r)$ 
controls not only the boundary condition at infinity, but also the 
possibility for the string to turn around and come back to the brane at 
infinity.

\subsection{Energy and Separation of the $Q\bar{Q}$ pair.}
We now follow the standard treatment for Wilson loops summarized in 
\cite{Sonnenschein:1999if}. If our probe-string hangs from infinity, turns 
around at a point $\r_0$ as described above and goes back to the 
$D$-brane at infinity, we can 
then compute gauge theory quantities, like the  separation between the two 
ends of the string, which can be thought  of as  the separation between a quark-antiquark 
pair living on the $D$-brane and coupled to the end-points of the string.
And we can compute the Energy of the pair of quarks, that we  associate with 
the length of the string (computed along its path in the bulk).
 Both of these quantities will be functions of 
the turning point $\r_0$.

The standard expressions that we will use can be derived easily. Indeed, 
for the $Q\bar{Q}$ separation we only need to compute $\int dx$. To 
calculate the 
energy of the  $Q\bar{Q}$ pair, we compute the action of the string and 
substract the action of two `rods' that would fall from infinity to 
the end of the space \footnote{Notice that these `rods' need not be 
strings, it is just a way of renormalizing the infinite mass of the 
quarks.}. 
The results are \cite{Sonnenschein:1999if}, 
\bea & & L_{QQ}(\r_0)=2 
f(\r_0)\int_{\r_0}^{\infty}\frac{g(z)}{f(z)}\frac{dz}{\sqrt{f^2(z)- 
f^2(\r_0)}},\nonumber\\
& &  E_{QQ}(\r_0)=
f(\r_0) L_{QQ}(\r_0) 
+ 2\int_{\r_0}^{\infty}\frac{g(z)}{f(z)}\sqrt{f^2(z)-f^2(\r_0)} dz- 2 
\int_{\hat{\r}_0}^{\infty} g(z) dz\,.
\label{EL}
\eea
As discussed above, the constant $C$ defined around Eq.~(\ref{drdx2}) must 
be taken to be $C= f(\r_0)$. Using Eq.~(\ref{nnzz})  we can rewrite
\beq
L_{QQ}(\r_0)=2 \int_{\r_0}^{\infty} \frac{dz}{V_{eff} (z)}
\label{LVeff}
\eeq
As  discussed in section \ref{boundaryconds}, we must 
impose that for large values of the radial coordinate $V_{eff}$ diverges. 
This, however, is not enough to ensure that the integral above receives a 
{\it  finite} 
contribution from the upper end of the integral, we then have to require 
that for large radial coordinate $V_{eff}$ diverges {\it at least} as
\beq
V_{eff}\underset{\r\to\infty}{\sim} \r^{\beta}\,,
\label{zzz1}\eeq
with $\beta > 1$. Then, the quantity 
$L_{QQ}$ can be finite or infinite depending on the IR 
($\r\to \r_0 \equiv \hat{\r}_0$) 
asymptotics of the background, meaning that we consider the 
turning point at the end of the space at the lower end of the integral. 
If, 
expanding around the turning point $\r_0$, we have
\beq
V_{eff}\underset{\r\to\r_0}{\sim} (\r-\r_0)^{\gamma}\,,
\label{zzz}\eeq
then the separation of the quark pair
is infinite for $\gamma \geq 1 $ and finite for $\gamma<1$. We will see 
examples of both behaviors in the following sections.

Another reasonable condition that we could impose is that $\r_0$ is the 
first 
zero of $V_{eff}$ for which the string has 
positive convexity (a minimum). This  can be easily expressed as a 
condition on 
the background functions. In fact, let us assume that near $\r_0$ the 
effective potential 
in Eq.~(\ref{nnzz}) behaves as $V_{eff}= \kappa 
(\rho-\rho_0)^{\gamma}$. Then, in a
neighbourhood of $\r_0$, we have
\bea
& & \frac{d\r}{dx}=\pm V_{eff}=\pm\left(\kappa (\r-\r_0)^\gamma\right) +\mathcal{O}(\r-\r_0)^{\gamma+1}\nonumber\\
& & \frac{d^2\r}{dx^2}= \frac{d}{dx}(\frac{d\r}{dx})=\pm\frac{d 
V_{eff}(\r)}{d\r} \frac{d\r}{dx}= V_{eff}(\r) \frac{d 
V_{eff}(\r)}{d\r}=\kappa^2 \gamma (\r-\r_0)^{2\gamma -1}+\mathcal{O}(\r-\r_0)^{2\gamma},
\label{nnhh}
\eea
 by iteration and induction we obtain (we choose only one of the branches of Eq.~(\ref{Eq:eqofmot})),
\beq
\frac{d^n\r}{dx^n}= (V_{eff}(\r) \frac{d}{d\r})^n \r
= \kappa^n \Pi_{j=0}^{n-1} \Big[ j(\gamma-1) +1\Big]
(\r -\r_0)^{(\gamma -1) n + 1}+\mathcal{O}(\r-\r_0)^{(\gamma -1) n + 2}.
\label{fff}\eeq
In order to have a minimum we impose that the first non vanishing 
derivative at $\r=\r_0$ is an even derivative (its value has to 
be positive).
So, for even $n$ we need (this result will not depend on the choice of branches above),
\beq
\gamma = 1 -\frac{1}{n},
\label{janzz}
\eeq
from this it follows that $\gamma$ cannot be bigger than one. 
This 
means that the only possibility 
of obtaining a string that can be stretched 
up to infinite distance will come from the case $\gamma=1$ 
(contrast this with the result of \cite{Kinar:1998vq}). 
The 
solutions with infinite length will have all even derivatives vanishing at 
$\r_0$. 
Had we considered non-integer corrections to the leading term in 
Eq.~(\ref{nnhh}), so that 
near $\r_0$ the function  
\beq
V_{eff}= \kappa_1
(\rho-\rho_0)^{\gamma}+\kappa_2
(\rho-\rho_0)^{\gamma+\epsilon}
\eeq
for very small values of $\epsilon$ we would have had a formula like the 
one in 
Eq.~(\ref{fff}) where the exponent is the same, but the coefficient 
changes. 
For large values of $\epsilon$, we have, to leading order, the same result as 
in Eq.~(\ref{fff}). The reasoning given above applies.

These constraints ensure 
that there exists a unique trajectory for the string as function of 
$\r_0$.
It should be also noticed that the profile of the string is not analytic function and in particular in $\r_0$ it turns out to be a $\mathcal{C}^2$ function.
\subsubsection{Some exact results.}
 Next, let us derive an  expression of the energy $E_{QQ}$ as 
function of 
the inter-quark separation $L_{QQ}$. 
For this purpose, it is useful to introduce the function
$K[x]=\frac{1}{\sqrt{x^2-1}}$. The expression for $L_{QQ}(\r_0)$ in
Eq.~(\ref{EL}) reads\footnote{Some of these expressions have been 
derived in past 
collaborations with Angel Paredes.}
\beq
L_{QQ}(\r_0)=\lim_{\r_1\to\infty}2 \int_{\r_0}^{\r_1}\frac{g(z)}{f(z)}K\left[\frac{f(z)}{f(\r_0)}\right]{dz}\,.
\eeq
Performing the derivative
\beq
\frac{dL_{QQ}}{d\r_0} = - 2 \frac{g(\r_0)}{f(\r_0)} K[1] + 
\lim_{\r_1\to\infty} 2 
\int_{\r_0}^{\r_1}\frac{g(z)}{f(z)}\partial_{\r_0}K\left[\frac{f(z)}{f(\r_0)}\right]
\eeq
Now, we use the following identity,
\beq
\partial_{\r_0}K\left[\frac{f(z)}{f(\r_0)}\right]= 
- \partial_{z}K\left[\frac{f(z)}{f(\r_0)}\right]\,
\left( \frac{f(z) f'(\r_0)}{f'(z) f(\r_0)}\right)
\eeq
integrating by parts and after some algebra, we get
\beq
\frac{dL_{QQ}}{d\r_0} = - \lim_{\r_1\to\infty}2\partial_{\r_0}\log
f(\r_0) \left\{
\frac{g(\r_1)}{f'(\r_1)}K\left[\frac{f(\r_1)}{f(\r_0)}\right]
- \int_{\r_0}^{\r_1}dz~K\left[\frac{f(z)}{f(\r_0)}\right] 
\partial_z\left(\frac{g(z)}{f'(z)}\right)  
\right\}\,,
\label{dldr0}
\eeq
or 
\beq
\frac{dL_{QQ}}{d\r_0} = 2 \partial_{\r_0}\log f(\r_0) 
\lim_{\r_1\to\infty}\left\{
\int_{\r_0}^{\r_1}dz~K\left[\frac{f(z)}{f(\r_0)}\right]
\partial_z\left(\frac{g(z)}{f'(z)}\right) 
- \frac{f(\r_1)}{f'(\r_1) V_{eff}(\r_1)} \right\}\,.
\label{cnzz}
\eeq
Similarly, we compute  the derivative for the energy function finding that
\beq
\frac{dE_{QQ}}{d\r_0}=  f(\r_0)\frac{dL_{QQ}}{d\r_0} \,,
\eeq
and  we get
\beq
\frac{dE_{QQ}}{dL_{QQ}}=f(\r_0)\,.
\label{dEdLfinal}\eeq
Notice that (after integration) Eq.~(\ref{dEdLfinal}) is an {\it exact} 
expression for 
$E_{QQ}$ in terms of $L_{QQ}$, the non triviality residing in the function $\r_0(L_{QQ})$, this expression was also derived in \cite{Brandhuber:1999jr}. Also, Eq.~(\ref{dEdLfinal}) yields the force between the quark pair.
It should be stressed that we are 
assuming that we will be able to invert 
$\r_0$ as function of $L_{QQ}$. Whenever this cannot be done, the solution 
would be valid only in the regions of monotonicity of $L_{QQ}(\r_0)$.

Let us comment briefly about the existence of cusps in our strings.
If near the lower point of the string (typically situated in the end
of the geometry) the quantity $V_{eff}(\r)$ diverges, then there will be a 
cusp in the shape of the string. This can be easily understood using 
Eq.~(\ref{drdxfinal}) and an expansion of $V_{eff}(\r)$ of the form 
discussed in Eq.~(\ref{Eq:eqofmotcompact}). If $V_{eff}(\r)\propto(\r-\r_0)^\gamma$ with $\gamma<0$, one can integrate the equation
\beq
\r-\r_0\propto \left[ (1-\gamma)\left|x-\frac{L_{QQ}}{2}\right| \right]^{\frac{1}{1-\gamma}},
\eeq
which implies that the shape of the string $\r(x)$ is not analytic at 
$\frac{L_{QQ}}{2}$.

\subsection{Leading and subleading behaviors. Inversion points}
Let us focus on the case that, near the end of the space (in the IR),
$V_{eff}\sim (\r-\r_0)$ (the string stretches up to 
infinite length $L_{QQ}(\hat{\r}_0)\to \infty$). 
The
subleading corrections to the Energy of the quark pair \cite{Kinar:1998vq} 
 read
\beq
E_{QQ}= f(\hat{\rho}_0) L_{QQ} +\kappa + O(e^{- |a| L_{QQ}} (\log L_{QQ})^b)\,.
\label{subleadingEnergy}
\eeq
This formula can be obtained as an expansion around $\r_0\to\hat{\r}_0$, 
or equivalently expanding Eq.~(\ref{dEdLfinal})  around 
$L_{QQ}\to\infty$.
Notice that no power-law corrections
or Luscher-terms appear in these corrections: they would appear 
if $\gamma>1$ in Eq.~(\ref{janzz}), something that we ruled out. 
Contrast this with the result of \cite{Kinar:1998vq}. 
Because  Eq.~(\ref{janzz}) and the discussion around it rely only on
the generic properties of $V_{eff}$, such power-law corrections can only
descend from higher-derivative corrections to Eq.~(\ref{ng}),
which cannot be repackaged into the form of Eq.~(\ref{drdxfinal}).

In the following sections, we will discuss these subleading behaviors in 
various examples and precisely find the coefficients that apply to each 
particular case.
Before proceeding, a few  comments are due  in order to avoid
ambiguities.

The partition function of the string is studied  by using two different expansions, in $\alpha'$ and $g_s$.
First, the Nambu-Goto action is characterized by the string tension,
and one can think of expanding any physical quantity (correlation function)
in powers of $\alpha^{\prime}$. This procedure is effectively equivalent to
quantizing the (particle-like) excitations of the string, and in this sense $\alpha^{\prime}$ corrections
can be associated with loops of the dynamics of the string modes.
One can also rephrase the results in terms of a classical effective theory, in 
which the $\alpha^{\prime}$ corrections are encoded in higher derivative corrections
to the Nambu-Goto classical action.
By doing so, one can associate the resulting $E_{QQ}$ to the expectation value of
a Wilson loop in the dual theory, and hence interpret the generalization of Eq.~(\ref{subleadingEnergy})
in terms of  the actual quantum corrections to the quark-antiquark potential at large 
distance, for a confining string, as done for example in~\cite{Aharony:2009gg}.
As explained above, this procedure yields power-law corrections to the leading-order result
 $E_{QQ}\propto L_{QQ}$.
This is {\it not} what we are doing in this paper.
All the results we obtain are at the leading-order in $\alpha^{\prime}$,
the calculations being  completely classical  and based on the Nambu-Goto 
action.
In particular, this explains why we do not find a Luscher term.

We are going to truncate at the leading-order also the expansion in $g_s$,
which controls processes where the string breaks. 
This is also done in~\cite{Aharony:2009gg}, where the quantum corrections 
that are computed are the ones in $\alpha^{\prime}$, but the whole analysis 
treats the string itself as effectively free. Not only quantum corrections
related to the $g_s$ expansion, but the whole many-body nature of 
string-theory is neglected in this way.
This is a very important limitation: strings that can stretch up to infinite separation for the quark pair
can be treated in this way, while theories in which fragmentation and hadronization
take place via string-breaking are accessible to our 
approach only up to a 
scale smaller than the scale of breaking itself (real world QCD, if it admits a 
string dual description, should fall into this class 
in which $g_s$ corrections must be included 
and actually dominate the dynamics from the scale of breaking and below). 

We will see  in the following 
examples in which Eq.~(\ref{zzz}) holds with $\gamma<1$. 
This will yield a finite value for the maximal displacement in the Minkowski directions
between the quark-antiquark pair.
Studying the large-$L_{QQ}$ limit would necessarily require 
the inclusion both of quantum effects in the $\alpha^{\prime}$ expansion,
but also the effect of string breaking, which might be decoupled in the
large quark-mass limit.

Another important comment is the following:
if the function $L_{QQ}(\r_0)$ is not monotonic, then it is not invertible, 
and $\r_0(L_{QQ})$ is multivalued.
If this is the case, it will happen that for some given 
$L_{QQ}$ we can find that also $E_{QQ}(L_{QQ})$ is multivalued.
Among all the possible values of $E_{QQ}$ for a given $L_{QQ}$ 
(all of which satisfy the equations of motion)
we will  refer to the lowest one as the stable solution, 
while the others will represent either excited or unstable states.
The Van der Waals liquid-gas system provides a nice realization of these 
excited and unstable states, and we report its treatment in 
Appendix~\ref{vanderwaals}.

The existence of inversion points (or extrema of $L_{QQ}(\r_0)$)
implies that the first derivative $\frac{d L_{QQ}}{d\r_0}$  vanishes. 
Using the expression in Eq.~(\ref{cnzz}),
\beq
\lim_{\r_1\to\infty}\left\{    
\int_{\r_0}^{\r_1}dz~K\left[\frac{f(z)}{f(\r_0)}\right]
\partial_z\left(\frac{g(z)}{f'(z)}\right) 
- \frac{f(\r_1)}{f'(\r_1) V_{eff}(\r_1)} \right\}=0.
\label{rara}
\eeq
If it happens that $\frac{f(\r_1)}{f'(\r_1) V_{eff}(\r_1)}\to 0$ for large 
values of $\r_1$, then a necessary condition for the existence of turning 
points is that the integral in Eq.~(\ref{cnzz}) vanishes, or more simply 
that 
the integrand changes sign. 
Notice, nevertheless, that this is a not a criterium of practical application except for some particular easy backgrounds as we will see in section \ref{flavoredbackgroundsxx}.\\
To close this section let us stress that given the constraints we have specified along this section, the shape of the string that we are proposing as solution is the only allowed shape, since there can be only one minimun of $\r(x)$.

%
\section{Some well-understood examples}\label{examplesunderstood}
In this section we illustrate the points of the previous section in 
a set of very well known examples. All of them are 
string-theory backgrounds that are simple enough that the dynamics of the
probe string can be treated analytically.
We will write the 10-dimensional background in string frame.

\subsection{The case of $AdS_5 \times S^5$}
This is certainly the  main example, where the conjecture was 
originally  proposed~\cite{Maldacena:1997re}
and the ideas for Wilson loops 
described in the 
introduction first developed~\cite{Maldacena:1998im}.
After a rescaling of the radial coordinate, the metric reads,
\beq
ds^2= \alpha' \Big[\frac{\r^2}{R^2}dx_{1,3}^2 +\frac{R^2}{\r^2}d\r^2 + R^2 
d\Omega_5^2   \Big]\,,
\eeq
where $R^2=\sqrt{4\pi g_s N_c}$ is dimensionless, $\r$ has dimensions of inverse-length,
 and the constant $\alpha^{\prime}$ in front of the metric ensures that when we compute the functions $f,g$ 
defined below Eq.~(\ref{zxzx}) and plug into Eq.~(\ref{ng}) the factors of 
$\alpha'$ will cancel (this will happen in the examples discussed below 
also). 
Hence we have,
\beq \begin{cases}
g(\r)^2=\alpha^{\prime\,2}&\\
f(\r)^2=\alpha^{\prime\,2}\frac{\r^4}{R^4}&\\
C^2=f(\r_0)^2=\alpha^{\prime\,2}\frac{\r_0^4}{R^4}&
\end{cases}
V_{eff}=\frac{\r^2}{\r_0^2 R^2}\sqrt{\r^4-\r_0^4}
\eeq
One can check that the functions of the background respect all the 
constraints we imposed in section \ref{general} regarding the boundary 
conditions 
and convexity.
We can exactly integrate $L_{QQ}(\r_0)$
\beq
L_{QQ}(\r_0)=2 \int_{\r_0}^{\infty}d\r\frac{R^2\r_0^2 }{\r^2 
\sqrt{\r^4-\r_0^4}}=(2 \pi)^{\frac{3}{2}} \frac{R^2}{\r_0 \Gamma (\frac{1}{4})^2}
\eeq
Since it is possible to invert the relation we can 
then write $\r_0(L_{QQ})$, and using Eq.~(\ref{dEdLfinal})
\beq
\frac{dE_{QQ}}{dL_{QQ}}=\frac{(2\pi)^3 R^2}{L_{QQ}^2 \Gamma (\frac{1}{4})^4} \Rightarrow E_{QQ}(L_{QQ})=-\frac{(2\pi)^3 R^2}{L_{QQ} \Gamma (\frac{1}{4})^4}
\label{ads5el}
\eeq
in agreement with the result of \cite{Maldacena:1998im}.
Of course, the separation for the quark pair diverges for $\rho_0\to 0$ 
(the end of the space). 
Also notice that in this case, the expression of $L_{QQ}(\r_0)$ is 
invertible. There are no corrections to Eq.~(\ref{ads5el}). 

A few words of comment might be useful to a reader who is not familiar with 
this set-up. The Yang-Mills coupling of the dual ${\cal N}=4$ field theory
is defined as $g_{YM}^2\equiv 2\pi g_s$, so that the (dimensionless) curvature 
$R^4$ is proportional to the 't Hooft coupling in the CFT. The supergravity
approximation holds when $R^4\gg 1$.
\subsection{Witten-Sakai-Sugimoto Model}
This model, based on  $D4$ branes wrapped on a  circle with SUSY breaking 
periodicity conditions \cite{Witten:1998zw} received a great deal of 
attention thanks to the observation by Sakai and Sugimoto~\cite{Sakai:2004cn}
that the introduction of  $N_f\ll N_c$
flavor  $D8$ branes as probes allows to 
construct a model where a peculiar realization of chiral symmetry is spontaneously 
broken.
The metric  reads
\beq
\frac{ds^2}{\alpha'}= \left(\frac{\r}{R}\right)^{3/2}  \left[\frac{}{}dx_{1,3}^2 + \hat{F}(\r) 
dx_4^2\right] 
+\left(\frac{R}{\r}\right)^{3/2} \left[\frac{}{}\frac{d\r^2}{\hat{F}(\r)} +\r^2 d\Omega_4^2\right]\,,
\eeq
where $x_4$ is the coordinate on the circle, $R^3=\pi g_s \sqrt{\alpha^{\prime}}N_c$, 
and $\hat{F}(\r)=\frac{\r^3-\Lambda^3}{\r^3}$. 
Notice that in this case the gauge coupling $g_{YM}^2\equiv (2\pi)^2 g_s \sqrt{\alpha^{\prime}}$
is dimensionful, so that $\r$ has dimensions of inverse-length, as in the previous subsection,
and so does $\Lambda$.
The relevant functions are,
\beq
\begin{cases}
f^2(\r)=\alpha^{\prime\,2}\frac{\r^3}{R^3}&\\
g^2(\r)=\alpha^{\prime\,2}\frac{\r^3}{\r^3-\Lambda^3}&\\
C^2=f(\r_0)^2=\alpha^{\prime\,2}\frac{\r_0^3}{R^3}&
\end{cases}
V_{eff}=\sqrt{\frac{(\r^3-\Lambda^3)(\r^3-\r_0^3)}{(\r_0 R)^3}}
\eeq
hence,
\beq
L_{QQ}=2(\r_0 R)^{3/2} \int_{\r_0}^{\infty} 
\frac{d\r}{\sqrt{(\r^3-\r_0^3)(\r^3- 
\Lambda^3)}}
\label{inetgralssw}
\eeq
One can explicitly perform the integral above, but the  result, in terms 
of 
special functions, is not very 
illuminating. 
To get a better handle on the underlying dynamics, we consider the case in 
which $\rho_0=\Lambda$. In this case, we get
\beq
L_{QQ}(\rho_0)=-\frac{2 \sqrt{R^3 \r_0^3}}{6\r_0^2}
\left.\left[\sqrt{12}\arctan\left(\frac{2\r+\r_0}{\sqrt{3}\r_0}\right)  +\log\left(\frac{\r^2+\r\r_0 +\r_0^2}{(\r-\r_0)^2}\right)
\right] \right|_{\r_0}^{\infty}
\eeq
that we can see diverges logarithmically for $\r=\rho_0$. So, the string 
here again, has infinite length as is expected in a dual of a  confining 
field theory. If $\r_0>\Lambda$ then the separation 
of the pair is computed
from the integral of Eq.~(\ref{inetgralssw}) and turns out to be finite.

On the other hand,  it is possible to iteratively invert the relation 
between $L_{QQ}$ and $\r_0$ for $\r_0\sim\Lambda$ and hence for 
$L_{QQ}\to\infty$ \beq
\r_0(L_{QQ})=\Lambda + 4 \sqrt{3} e^{-\frac{\pi}{2 \sqrt{3}} - 
\frac{3 L_{QQ} \sqrt{\Lambda}}{4 R^{\frac{3}{2}}}} \Lambda+ \dots
\eeq
and from Eq.~(\ref{dEdLfinal}) we get for the Energy of the pair $E_{QQ}$ 
in terms 
of the separation $L_{QQ}$,
\beq
E_{QQ}(L_{QQ})\underset{L_{QQ}\to\infty}{=} 
L_{QQ} \left(\frac{\Lambda}{R}\right)^\frac{3}{2}- O( e^{-\frac{\pi}{2 
\sqrt{3}} - \frac{3 L_{QQ} \sqrt{\Lambda}}{4 R^{\frac{3}{2}}}} \Lambda) + \dots
\eeq
\subsection{D5 branes on $S^2$}\label{d5ons2xx}
In this example the system consists of $N_c$ $D5$ branes that wrap a two 
cycle inside the resolved conifold preserving four 
supercharges (${\cal N}=1$ in four dimensions)~\cite{Maldacena:2000yy}. After a 
geometrical transition takes place, a  
background dual to this field theory contains a metric, dilaton and RR 
three form. In this case $g_{YM}^2\equiv (2\pi)^3g_s\alpha^{\prime}$.
We quote  only the part of the metric relevant to this 
computation (after a rescaling of the Minkowski coordinates is done)
\beq
\frac{ds^2}{g_s\alpha' N_c}= e^{\phi}\Big[   dx_{1,3}^2 + d\r^2 + 
ds_{int}^2\Big]\,.
\eeq
The relevant functions read,
\beq
\begin{cases}
g(\r)= f(\r)=e^{\phi}\,g_s\alpha' N_c&\\
e^{2\phi}=\frac{e^{2\phi_0}\sinh(2\r)}{2 
\sqrt{\r \coth(2 \r) - \frac{\r^2}{\sinh(2 \r)^2}-\frac{1}{4}}}
\end{cases}
V_{eff}=e^{-\phi(\r_0)}\sqrt{e^{2\phi(\r)} - e^{2\phi(\r_0)}}
\eeq
The integral defining the separation of the quark pair cannot be evaluated 
explicitly, but we can check that
the upper limit of the integral gives a finite contribution (it goes as 
$\int^{\infty} \r^{\frac{1}{4}} e^{-\r} d\r$), while from the 
lower extremun of the string  reaches the end 
of the space ($\rho_0\to 0$) we get
\beq
L_{QQ}\sim \int_{\r_0 =0} \frac{d\r}{\sqrt{e^{2\phi_0}  \r^2+ ...}}\sim 
\lim_{\rho_0 \to 0}\log(\rho_0)\to\infty.
\eeq
indicating that (like in the Witten-Sakai-Sugimoto  example discussed 
above ) 
strings that reach the end of the space correspond to an 
infinite separation between the quark pair, which in turn indicates the 
absence of screening (expected from the QFT that contains only fields 
with zero N-ality).

We compute the subleading terms in the Energy of the pair in terms of 
the separation to obtain,
\beq
E_{QQ}= e^{\phi(0)}L_{QQ} + O(e^{-\frac{2\sqrt{2}}{3} L_{QQ}})
\eeq
as above a clear sign of a confining dual QFT.
\subsection{Klebanov-Strassler Model}
Certainly,  the Klebanov-Strassler model
is the cleanest example for a dual to a  four-dimensional field theory
that confines in the IR and approaches  a 
conformal point in the UV (modulo important subtleties)~\cite{Klebanov:2000hb}. 
Here again, we will quote only the relevant part 
of the metric,
\beq
ds^2= h(\r)^{-1/2}dx_{1,3}^2 + 
h(\r)^{1/2}\epsilon^{4/3}\frac{d\r^2}{6K(\r)^2} + ds_{int}^2
\eeq
with
\beq
K^3(\r)=\frac{\sinh 2\r  -2\r}{2\sinh^3(\r)},\;\; h(\r)=2^{2/3}
(g_s \alpha' N_c)^2 \epsilon^{-8/3}
\int_{\r}^{\infty}\frac{x\coth x -1}{\sinh^2 x}(\sinh 
2x  -2x)^{1/3}.
\eeq
The functions $f,g,V_{eff}$ read,
\beq
f^2=h(\r)^{-1},\;\; g^2= \frac{\epsilon^{4/3}}{6K(\r)^2},\;\; V_{eff}=
\frac{\sqrt{6 h(\rho_0)}K(\r)}{\sqrt{h(\r) 
\epsilon^{4/3}}}\sqrt{h(\r)^{-1} - h(\r_0)^{-1}}
\eeq
In this example again, using the asymptotics for the various functions, it 
can be checked that when the string approaches the end of the space 
$\rho_0 =0$ the separation of the quark pair diverges logarithmically, as 
in the two previous examples.
Also, the energy of the pair reads
\beq
E_{QQ}= f(0) L_{QQ} + O(e^{-\frac{2 \epsilon^{2/3}}{\sqrt{3} a_0} L_{QQ}})
\eeq
with $a_0= \frac{h(0)}{(g_s\alpha' N_c \epsilon^{-4/3})^2}=1.1398..$.
As above, it is clear that the model shows confining behavior.
\section{Walking solutions in the $D5$ system, 
unflavored.}\label{walkingsolutionsxx}

This section is devoted to the specific case of a class of solutions
to the $D5$ system that exhibit {\it walking} behavior in the IR,
in the sense that a suitably defined gauge coupling 
becomes almost constant in a finite range of energies~\cite{Nunez:2008wi}.
We remind the reader about the set-up, based 
on the geometry produced by stacking on top of each other $N_c$ 
$D5$-branes that wrap on a
$S^2$ inside a CY3-fold and then taking
the strongly-coupled limit of the gauge theory on this stack (that leaves us 
in the supergravity approximation for this system). 
We introduce the
class of solutions of interest and apply the formalism of the previous sections
to these solutions.
As we will see, these solutions do  confine, in the sense defined earlier on,
but also show a remarkable behavior in the walking energy region,
leading to a phenomenology resemblant a first order phase transition.
As a result, the leading-order, long-distance behavior of the quark-antiquark potential is
linear, but the coefficient is very different from what computed in
section~\ref{d5ons2xx}.

\subsection{General set-up.}

We start by recalling the basic definitions 
that yield the general class of backgrounds 
obtained from the $D5$ system, 
which includes~\cite{Maldacena:2000yy} as a very special case.
We start from the action of type-IIB
truncated to include only gravity, dilaton and the RR 3-form $F$:
\beq
S_{IIB}=\frac{1}{G_{10}}\int d^{10}x \sqrt{-g}\Big[
R-\frac{1}{2}(\partial\phi)^2 -\frac{e^{\phi}}{12}F_3^2  \Big],
\eeq

We define the $SU(2)$ left-invariant one forms as,
\bea\lab{su2}
\tilde{\w}_1\,=\, \cos\psi d\tilde\theta\,+\,\sin\psi\sin\tilde\theta
d\tilde\varphi\,\,,\,
\tilde{\w}_2\,=\,-\sin\psi d\tilde\theta\,+\,\cos\psi\sin\tilde\theta
d\tilde\varphi\,\,,\,
\tilde{\w}_3\,=\,d\psi\,+\,\cos\tilde\theta d\tilde\varphi\,\,.
\eea
and write an ansatz for the solution~\cite{Papadopoulos:2000gj}
 assuming that
the functions appearing in the background depend 
only the radial coordinate $\r$, but not on $x$ 
nor the 5 angles $\theta,\tilde{\theta},
\phi,\tilde{\phi},\psi$ (in string frame):
\bea
ds^2 &=& \alpha' g_s e^{ \phi(\rho)} \Big[\frac{dx_{1,3}^2}{\alpha' g_s} +
e^{2k(\rho)}d\rho^2
+ e^{2 h(\rho)}
(d\theta^2 + \sin^2\theta d\varphi^2) +\nonumber\\
&+&\frac{e^{2 g(\rho)}}{4}
\left((\tilde{\omega}_1+a(\rho)d\theta)^2
+ (\tilde{\omega}_2-a(\rho)\sin\theta d\varphi)^2\right)
 + \frac{e^{2 k(\rho)}}{4}
(\tilde{\omega}_3 + \cos\theta d\varphi)^2\Big], \nonumber\\
F_{3} &=&\frac{N_c}{4}\Bigg[-(\tilde{\omega}_1+b(\rho) d\theta)\wedge
(\tilde{\omega}_2-b(\rho) \sin\theta d\varphi)\wedge
(\tilde{\omega}_3 + \cos\theta d\varphi)+\nonumber\\
& & b'd\rho \wedge (-d\theta \wedge \tilde{\omega}_1  +
\sin\theta d\varphi
\wedge
\tilde{\omega}_2) + (1-b(\rho)^2) \sin\theta d\theta\wedge d\varphi \wedge
\tilde{\omega}_3\Bigg].
\label{nonabmetric424}
\eea

The system of BPS equations can be rearranged in a convenient form, by rewriting the
functions of the background in terms of a set of functions $P(\rho),
Q(\rho),Y(\rho), \tau(\rho), \sigma(\rho)$ as~\cite{HoyosBadajoz:2008fw}
\beq
4 e^{2h}=\frac{P^2-Q^2}{P\cosh\tau -Q}, \;\; e^{2g}= P\cosh\tau -Q,\;\;
e^{2k}= 4 Y,\;\; a=\frac{P\sinh\tau}{P\cosh\tau -Q},\;\; N_c b= \sigma.
\label{functions}
\eeq
Using these new variables, one can manipulate the BPS equations to obtain a
single decoupled second order equation for $P(\rho)$, while all other functions are
simply obtained from $P(\rho)$ as follows:
\bea
& & Q(\rho)=(Q_0+ N_c)\cosh\tau + N_c (2\rho \cosh\tau -1),\nonumber\\
& & \sinh\tau(\rho)=\frac{1}{\sinh(2\rho-2\hat{\rho_0})},\quad \cosh\t(\r)=\coth(2\r-2\hat{\r_0}),\nonumber\\
& & Y(\rho)=\frac{P'}{8},\nonumber\\
& & e^{4\phi}=\frac{e^{4\phi_o} \cosh(2\hat{\rho_0})^2}{(P^2-Q^2) Y
\sinh^2\tau},\nonumber\\
& & \sigma=\tanh\tau (Q+N_c)= \frac{(2N_c\rho + Q_o + N_c)}{\sinh(2\rho
-2\hat{\rho_0})}.
\label{BPSeqs}
\eea
The second order equation mentioned above reads,
\beq
P'' + P'\Big(\frac{P'+Q'}{P-Q} +\frac{P'-Q'}{P+Q} - 4 
\coth(2\rho-2\hat{\rho}_0)
\Big)=0.
\label{master}
\eeq

In the following we will fix the integration constant $Q_0=-N_c$, so that no singularity appears in the function $Q(\r)$. 
We also choose $\hat{\r}_o=0$ for notational 
convenience, together with $\alpha^{\prime}g_s=1$. For our purposes, it is also convenient to fix $8e^{4\phi_0}=1$.
With all of this, the functions we need for the probe string are:
\beqs
f^2(\r)&=&e^{2\phi}\,=\,\sqrt{\frac{\sinh^2(2\r)}{(P^2-Q^2)P^{\prime}}}\,,\\
g^2(\r)&=&\frac{1}{2}P^{\prime}f^2(\r)\,,\\
\label{Eq:VeffMN}
V_{eff}^2(\r)&=&\frac{2}{C^2P^{\prime}}\left(\sqrt{\frac{\sinh^2(2\r)}{(P^2-Q^2)P^{\prime}}}\,-\,C^2\right)\,.
\eeqs

\subsection{Walking solutions.}

The simplest and better understood solution of 
the system described in the previous subsection 
is defined by
\beqs
\label{Eq:MN}
\hat{P}&=&2 N_c \r\,.
\eeqs

It belongs to class I, and it has already been presented in section~\ref{d5ons2xx}, where is shown the study of the behaviour of the Wilson loop\footnote{see Appendix~\ref{uvasymptotics} for the definition of class I and II and for more details about the solutions to the equation for P.}.\\
For this background it is possible to show that
\beqs
f^2(0)&=&\frac{1}{N_c\sqrt{2N_c}}\,,
\eeqs
and that for $C^2=f^2(0)$, expanding around $\r\sim 0$
\beqs
V_{eff}^2(\r)&=&\frac{8\r^2}{9N_c}\,+\cdots\,.
\eeqs
In particular,  on the basis of what we already discussed around Eq.~(\ref{zzz})
this means that $L_{QQ}$ and $E_{QQ}$ {\it diverge} for $C^2\rightarrow f^2(0)$.

In~\cite{Nunez:2008wi}, it was observed that there exists a class of well-behaved solutions for which a suitably defined (four-dimensional) 
gauge coupling exhibits a {\it walking} regime, meaning 
that for a long range in the radial coordinate $\r_{IR}<\r<\r_{\ast}$ the 
running becomes very slow, and the gauge coupling effectively is constant. 
These solutions depend on two free parameters $c$ and $\alpha$,
and can be obtained recursively, by assuming $c$ large and expanding 
in powers of  $N_c/c $,  so that
\beqs\label{solution}
P(\r)&=&\sum_{n=0}^\infty c^{1-n} P_{1-n}.
\label{Eq:W}
\eeqs
with 
\be
P_1(\r)\equiv \left(\cos^3\a+\sin^3\a(\sinh(4\r)-4\r)\right)^{1/3}.
\ee
In particular, for $c/N_c\rightarrow +\infty$, the solution is very well approximated by 
$P\simeq c P_1$. 

The solutions of the form in Eq.~(\ref{Eq:W}) belong to class II in the language introduced in Appendix~\ref{uvasymptotics}.
This type of solution is not suited for the present study, because of the 
exponential behavior of $P$ and $P^{\prime}$ at large-$\r$, which is not compatible with 
the boundary conditions for the string in the UV as required in 
Eq.~(\ref{Eq:UVboundary}).
(Equivalently, the presence of 
high-dimensional operators dominating the dynamics in the far UV 
renders the study of the probes problematic. Analogous problems arise 
when studying the spectrum of excitations of the background~\cite{Elander:2009pk}
 and/or of  probe  fields~\cite{ENP}.)
However, we are mostly interested in what happens in the IR.
We are hence going to construct and study a different class of solutions,
which can be thought of as a generalization of Eq.~(\ref{Eq:W}),
with UV-asymptotics in class I. 
Such solutions can be expanded for small-$\r$, yielding,
\beqs
\label{Eq:Pexpan}
P(\r)=c_0+k_3 c_0\r^3+\frac{4 k_3 c_0\r^5}{5}
     -k_3^2c_0\r^6
+\frac{16 \left(2 c_0^2 k_3-5 k_3 N_c^2\right)\r^7}{105
   c_0}\,
+\cdots\,,
\eeqs
with $c_0$ and $k_3$ the two integration constants.
Notice how this expansion does not contain a term linear in $\r$. This means that
in the $c_0\rightarrow 0$ limit one does not trivially recover  Eq.~(\ref{Eq:MN}). 
 
In order to build numerically the solution, we start by expanding 
Eq.~(\ref{master}), by assuming that the solution can be written as
\beqs
P(\r)&=&\hat{P}(\r)\,+\,\varepsilon f(\r)\,,
\label{jlnzz}
\eeqs
and replacing in Eq.~(\ref{master}):
\beqs
0&=&F_0\,+\,\varepsilon F_1\,+\,{\cal O}(\varepsilon^2)\,.
\label{Eq:linear}
\eeqs
Because $\hat{P}$ is an exact solution, $F_0=0$.
Hence one finds a new equation, that now is linear:
\beqs
0&=&F_1\,=\,
\frac{8 \left(f(\r) (-\cosh (4 \r)+4 \r \sinh (4 \r)+1)-2 \r \sinh ^2(2 \r)
   f'(\r)\right)}{8 \r^2-4 \sinh (4 \r) \r+\cosh (4 \r)-1}+f''(\r)\,\\
   &\simeq&
   \frac{8 \left((1-4 \r) f(\r)+ \r f'(\r)\right)}{4
   \r-1}+f''(\r)
\eeqs
In the last step, we approximated the equation by assuming that $\r\gg 0$.
The resulting equation can be solved exactly, yielding, in terms of hypergeometric functions:
\beqs
f(\r)&=&e^{-4 \r} \sqrt{4 \r-1} c_1 U\left(\frac{5}{6},\frac{3}{2},6
   \r-\frac{3}{2}\right)+e^{-4 \r} \sqrt{4 \r-1} c_2
   L\left(-\frac{5}{6},\frac{1}{2},6 \r-\frac{3}{2}\right)\,.
\eeqs 
Asymptotically, neglecting power-law corrections, this means
\beqs
f(\r)&\simeq&c_1 e^{-4 \r} \,+c_2 e^{2\r}\,,
\eeqs
implying that consistency of the perturbative expansion in Eq.~(\ref{Eq:linear})
enforces the choice $c_2=0$.
Indeed, there are no asymptotic (in the UV) solutions
that behave as $e^{2\r}$. Allowing for $c_2\neq 0$
would imply that the solution is not a deformation of $\hat{P}$,
but rather that the expansion in Eq.~(\ref{jlnzz}) breaks down, 
and the solution (if regular) falls in class II.
The procedure of allowing for a small component $c_1\neq 0$ can be interpreted as the insertion of a small, relevant deformation, which does not affect the UV-asymptotics, but has very important physical effects in the IR.

\begin{figure}[thb]
\centerline{\includegraphics[width=12cm]{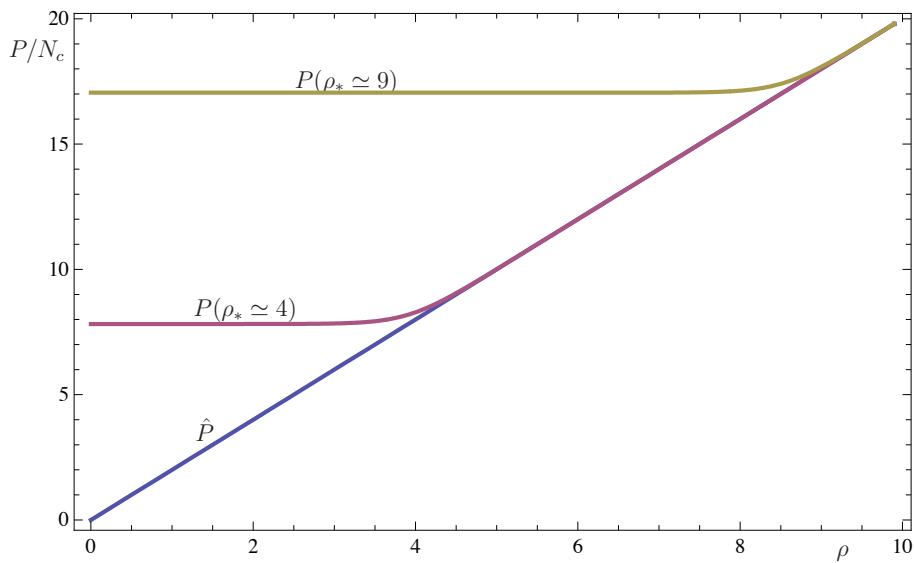}}
\caption{The numerical solutions for $P(\r)/N_c$ used in the analysis as an example, for  $N_c=100$. 
The three solutions correspond to the $\hat{P}$ case
with $\r_{\ast}=0$, and to two new numerical solutions with, respectively, $\r_{\ast}\simeq 4$ and
$\r_{\ast}\simeq 9$. The numerical solutions can be plotted up to $\r\simeq 150$, 
but in the following
we will truncate them at $\r_{1}=30$.}
\label{Fig:numericalP}
\end{figure}

\begin{figure}[htb]
\centerline{\includegraphics[width=7cm]{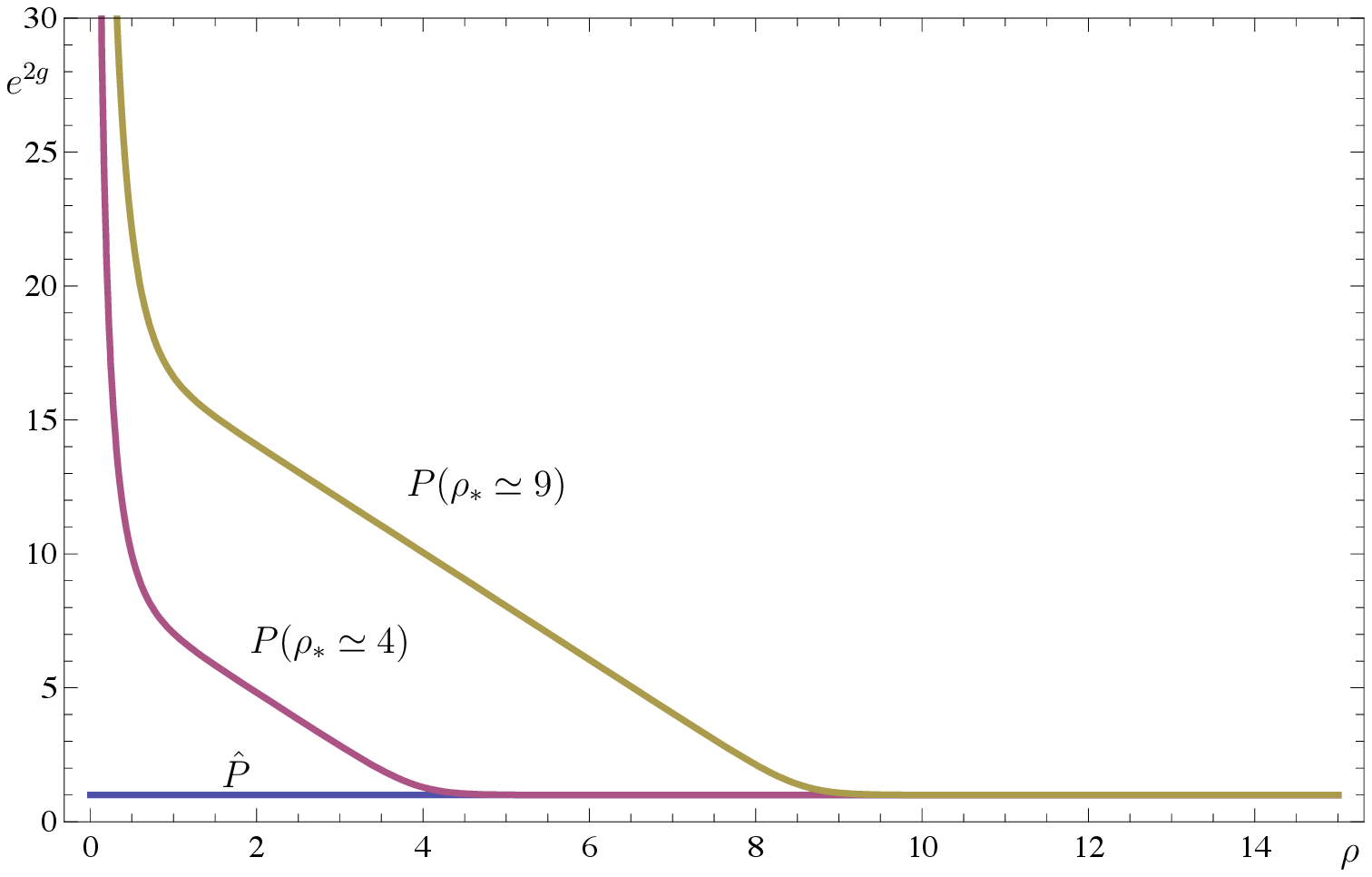}~~\includegraphics[width=7.1cm]{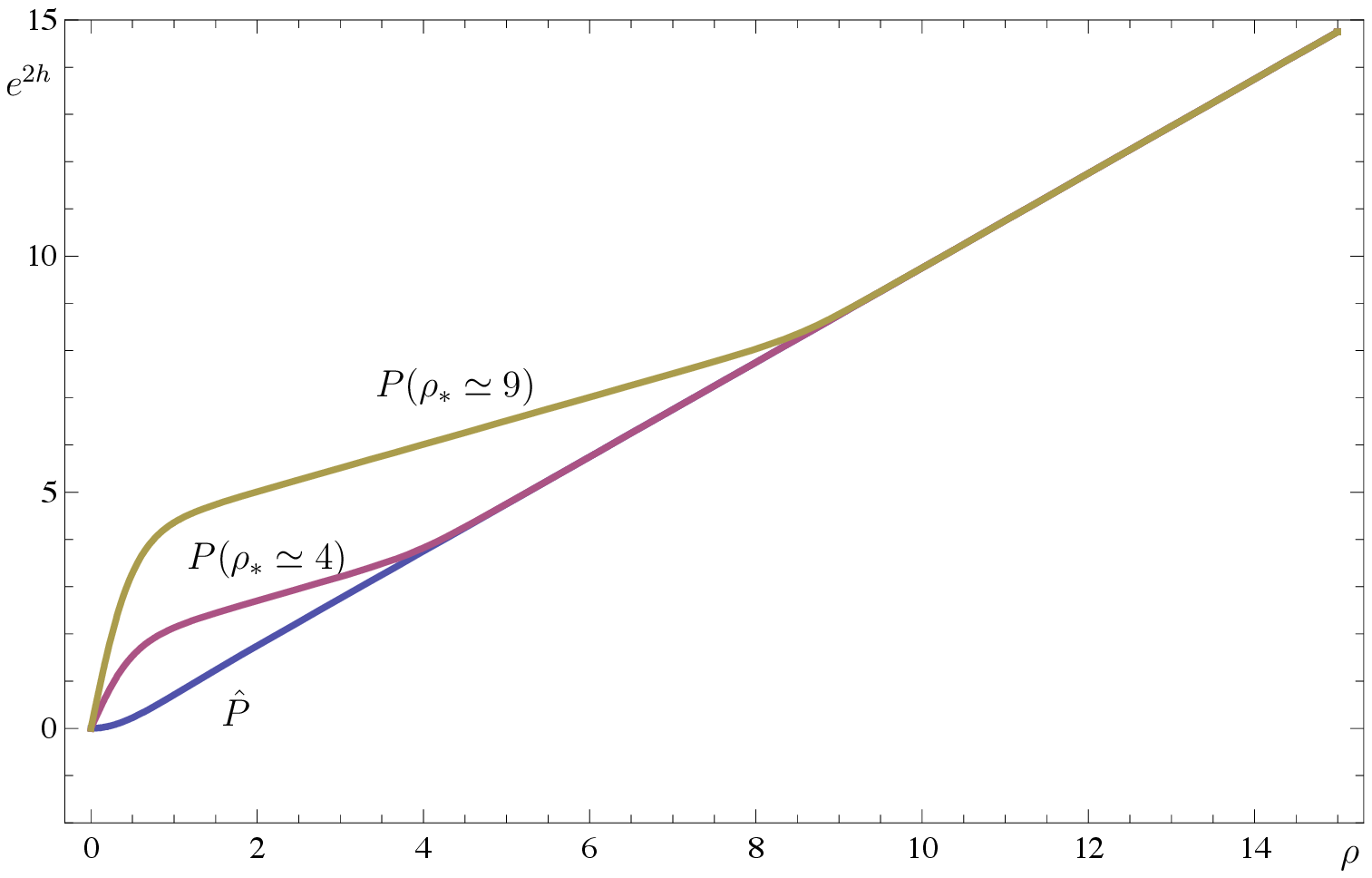}}
\centerline{\includegraphics[width=7.1cm]{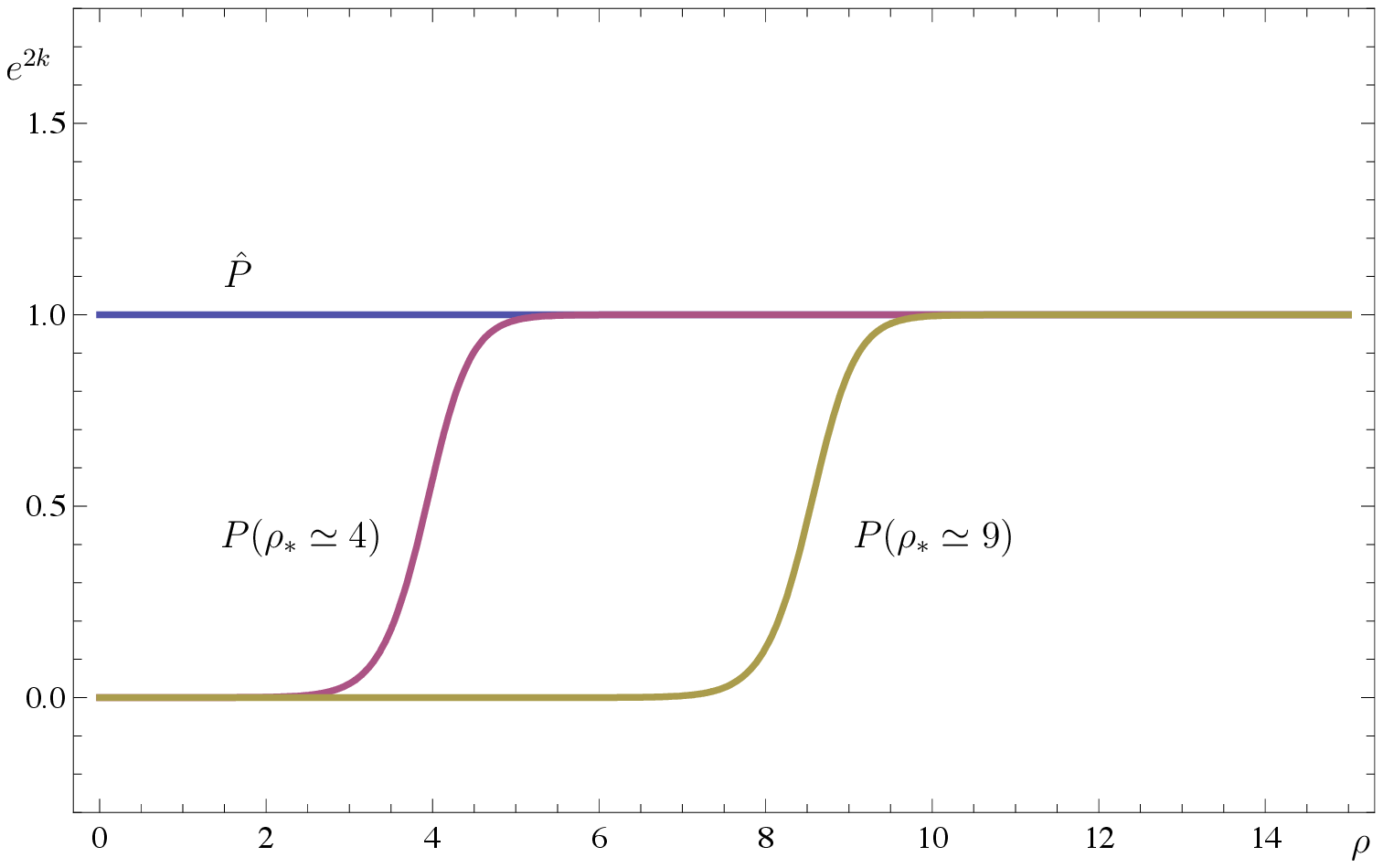}~~\includegraphics[width=7cm]{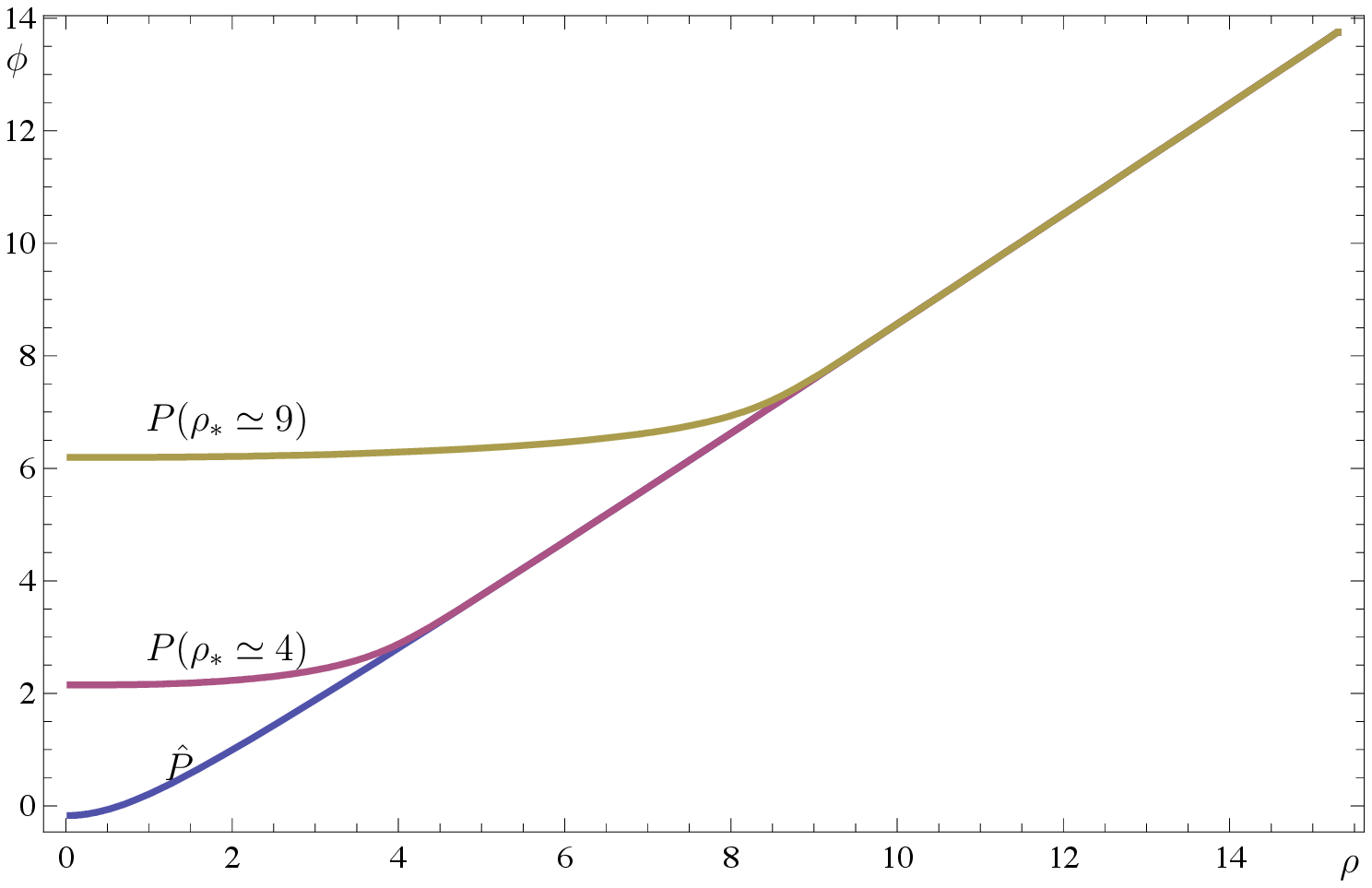}}
\caption{The functions $(e^{2g},e^{2h},e^{2k},\phi)$ appearing  in the metric
for the same solutions as in Fig.~\ref{Fig:numericalP}, computed rescaling 
$P\rightarrow P/N_c$, and $Q\rightarrow Q/N_c$.}
\label{Fig:background}
\end{figure}

We have now obtained an important result: at least up to large
values of $\r$, there exists a class of solutions that approach asymptotically the 
$\hat{P}$ solution.
We cannot prove that such solutions are well behaved all the way to $\r\rightarrow 0$.
However, we can use this result  in setting up the boundary conditions 
(at large-$\r$) and numerically solve Eq.~(\ref{master}) towards the IR.
By inspection, these solutions are precisely the ones we 
were looking for. They start deviating significantly from $\hat{P}$ 
below some $\r_{\ast}>0$, below which $P$ is approximately constant.
We plot in  Fig.~\ref{Fig:numericalP} two such solutions,
with $\r_{\ast}\simeq 4$ and  $\r_{\ast}\simeq 9$, together with 
the $\hat{P}$ solution for the same value of $N_c$.
We also plot in Fig.~\ref{Fig:background} the functions appearing in
metric $(e^{2g},e^{2h},e^{2k},\phi)$ for the same solutions.
Notice the behavior of $e^{2g}$ for $\r\rightarrow 0$, but also the fact that the dilaton $\phi$ is finite for $\r\rightarrow 0$.

\begin{figure}[hbt]
\centerline{\includegraphics[width=7cm]{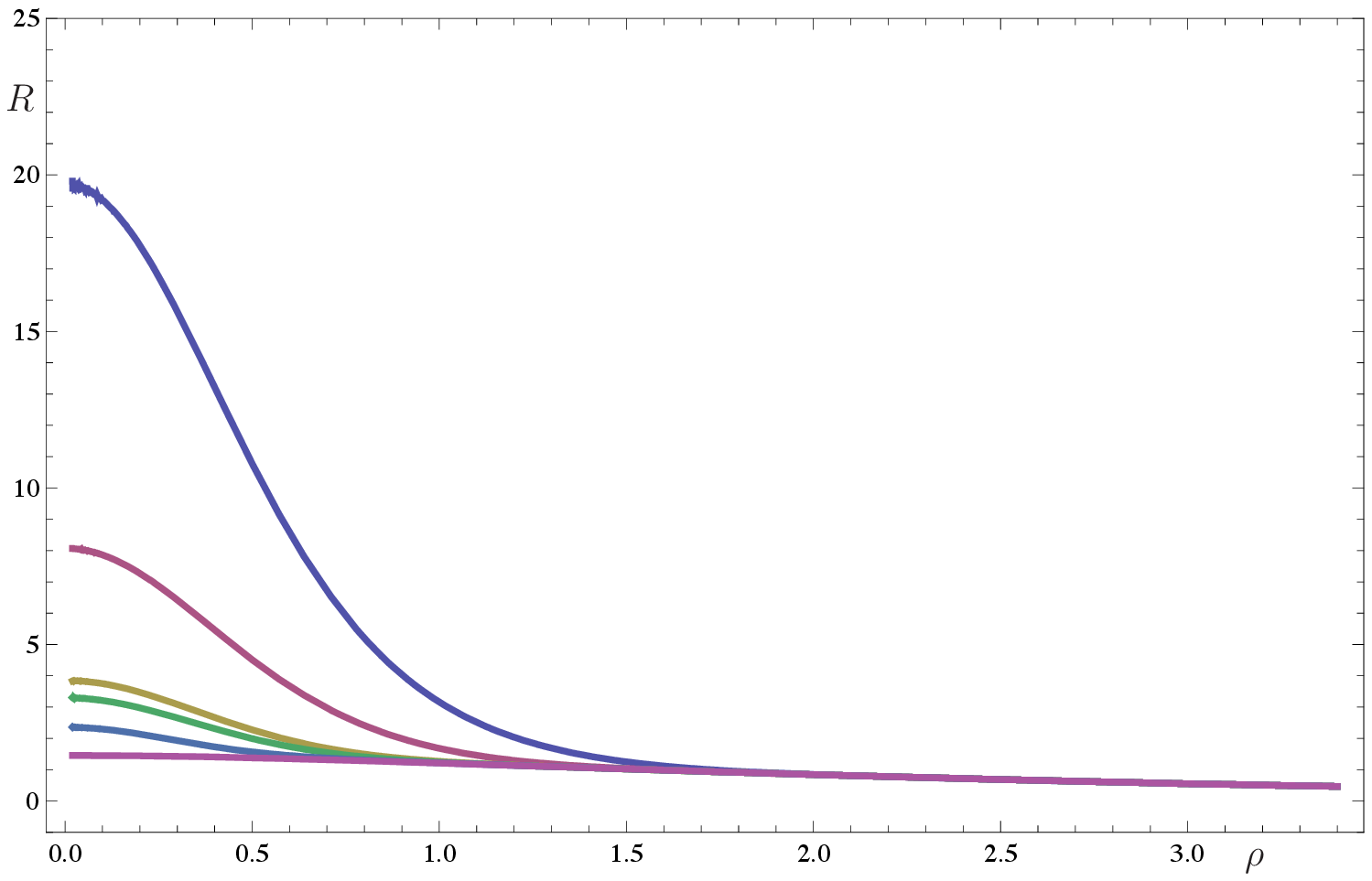}
\includegraphics[width=7.1cm]{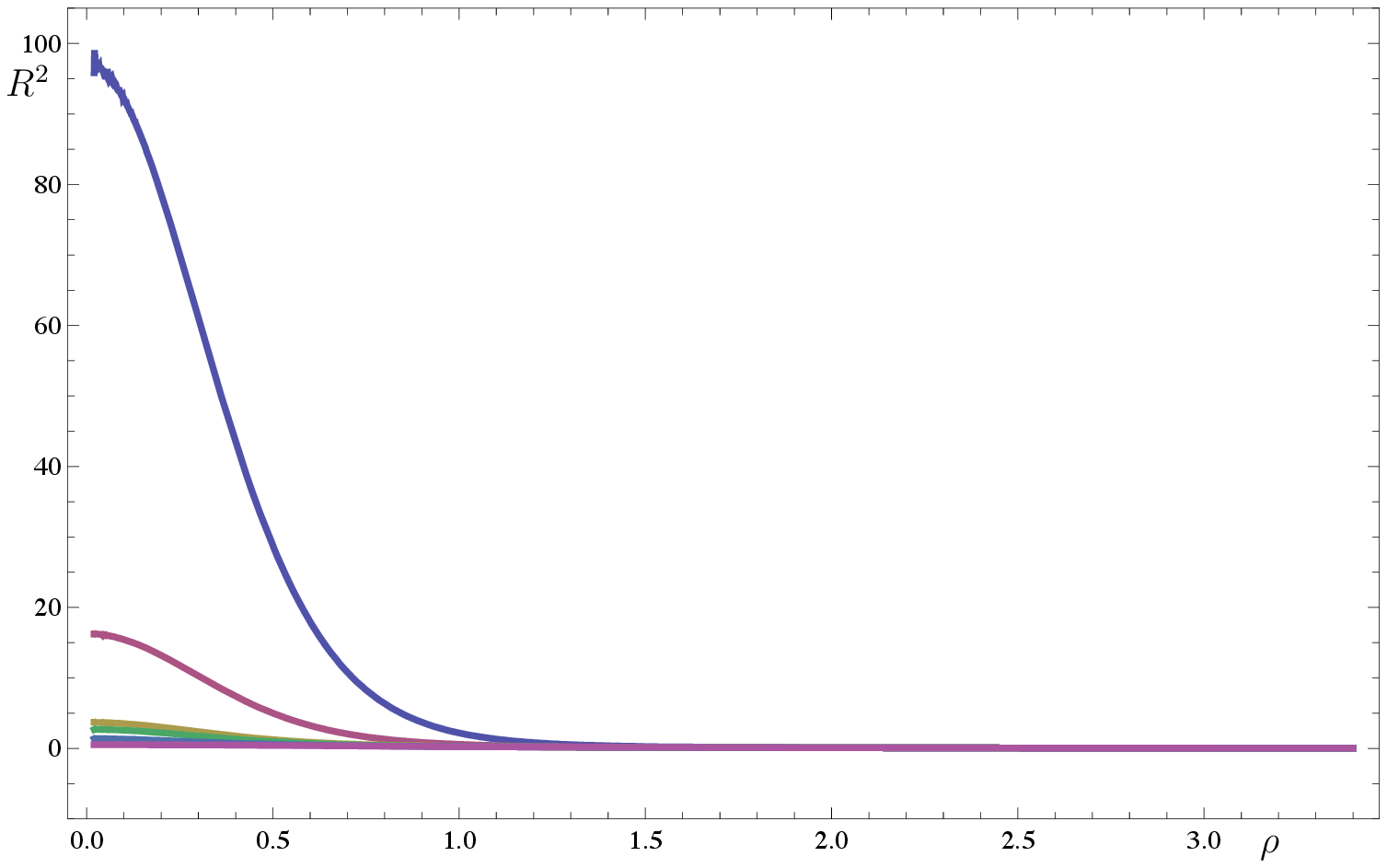}}
\caption{The (10-dimensional) Ricci scalar $R$ and the scalar $R^2\equiv R_{MN}R_{PQ}g^{MP}g^{NQ}$,
plotted as a function of the radial coordinate $\r$,
 for several numerical solutions in the class discussed in the body of the paper.
Each curve corresponds to a different value of $\r_{\ast}$. Notice that
both scalars are finite in the $\r\rightarrow 0$ limit. }
\label{Fig:Ricciscalar}
\end{figure}

A short digression is due at this point.
The reader should be aware that in the presented case we are likely working with a singular background, as the function $e^{2g}$, a warp factor inside the internal space, is divergent at $\rho=0$.\\
The problem of singularities in gauge-string duality has a long history. Surely it is better to work with non singular backgrounds, but it is possible to obtain interesting information also from singular manifolds. The procedure, developed along the years, to decide whether one is a dealing with a ``good singularity'' or not, consists in analyzing all the physical observables and showing that no track of the singularity can be found in them. One specific suggestion is to look at the behavior of the $g_{tt}$ component of the metric (in the present case, the dilaton) and make sure that it is finite. Many are in literature the examples of such a class of backgrounds. To corroborate the idea that our background falls in this class, we decide to investigate not only the value of the Wilson loop but also two invariants of the metric, namely the Ricci scalar and a suitable contraction of the Ricci tensor. As can be seen in Fig.~\ref{Fig:Ricciscalar} no trace of singularities can be found in these two observables for our background, in support of the idea that the backgrounds we are considering are indeed acceptable. 


\subsection{Probes: numerical study.}

\begin{figure}[htpb]
\centerline{\includegraphics[width=10.1cm]{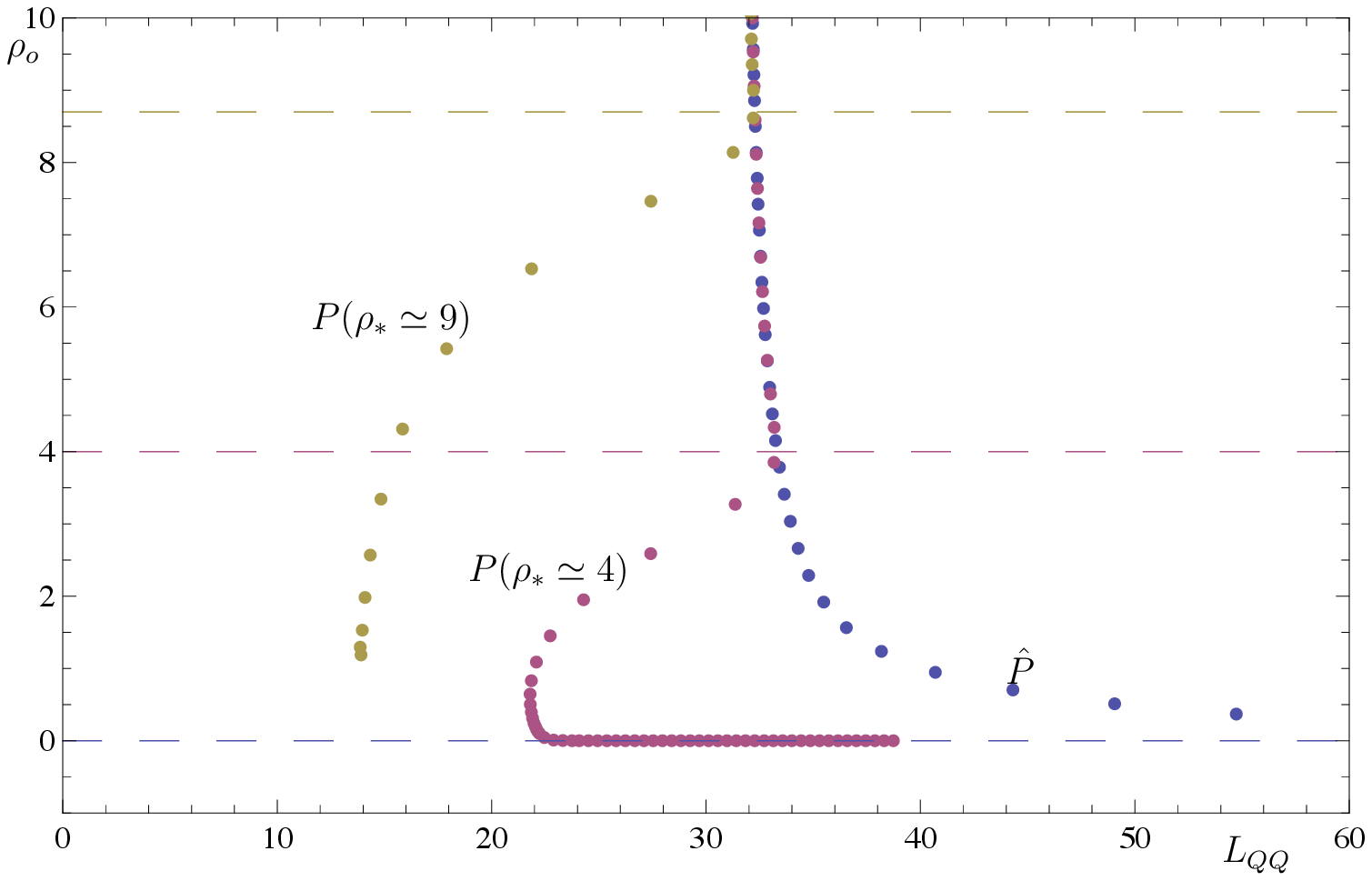}}
\centerline{\includegraphics[width=10.4cm]{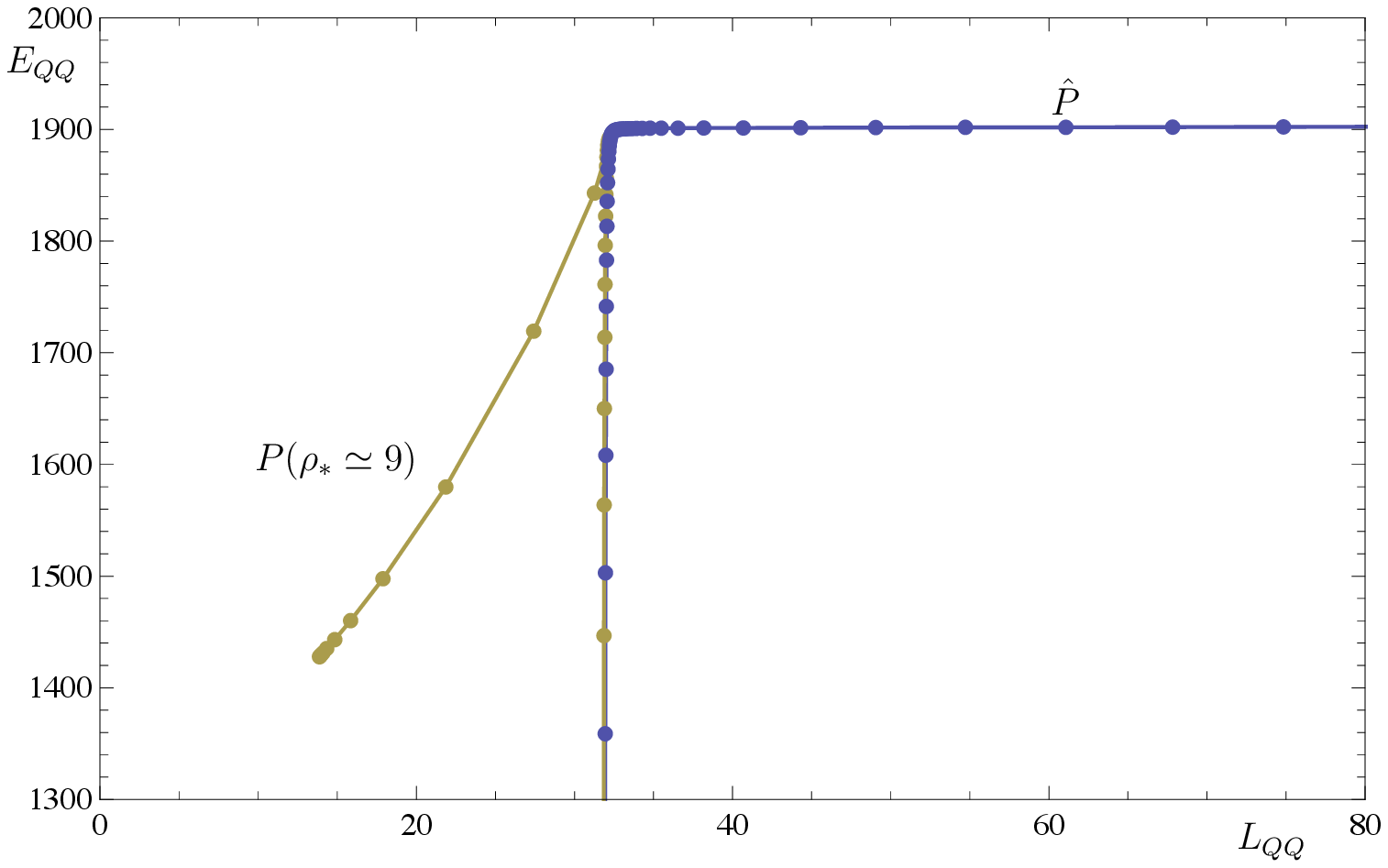}}
\centerline{\includegraphics[width=10.2cm]{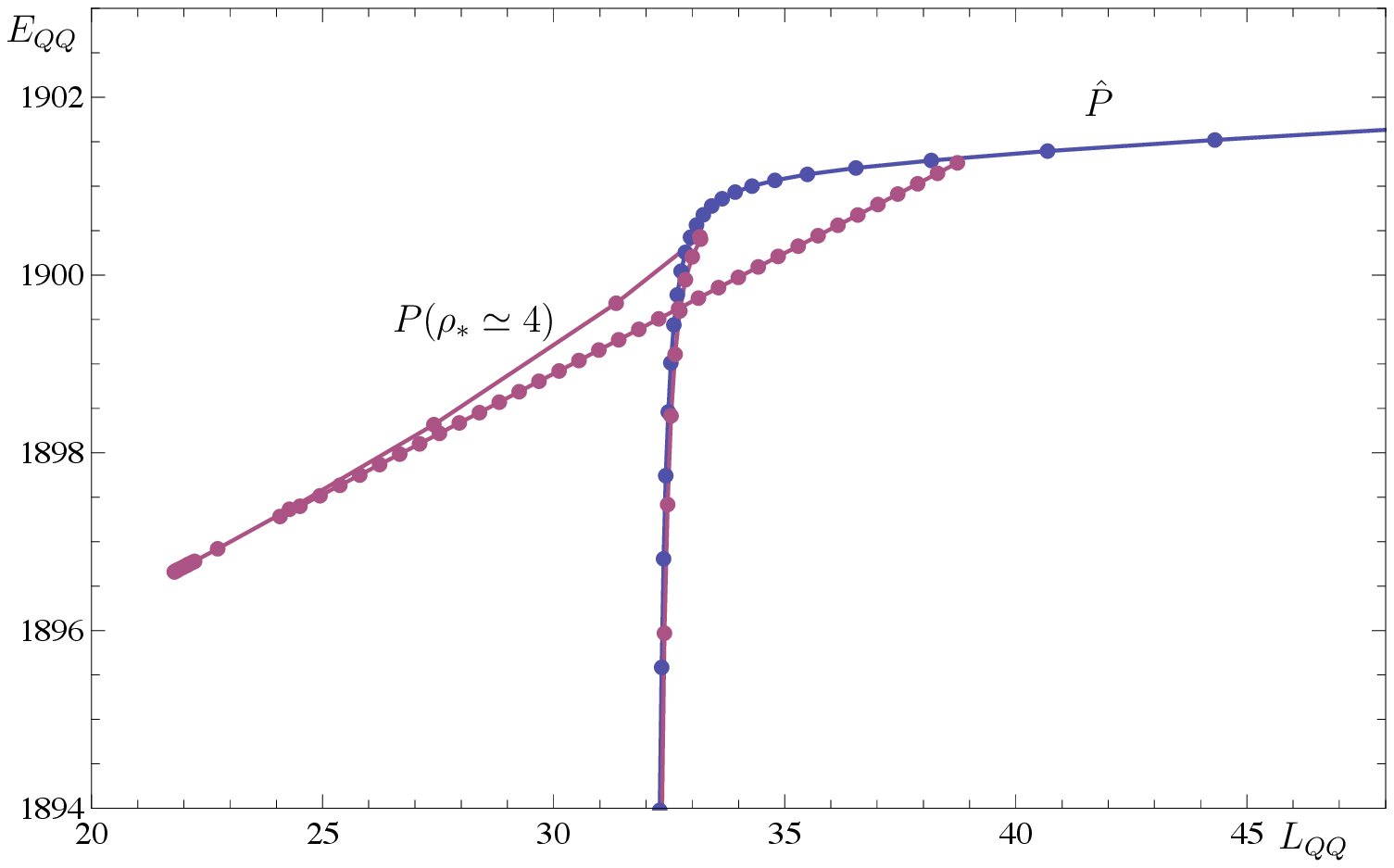}}
\caption{Upper panel, the radial coordinate $\r_0$ of the middle point of the string
as a function of $L_{QQ}$. Middle and lower panel, the energy $E_{QQ}$ as a function of the quark-antiquark separation $E_{QQ}(L_{QQ})$.
The three solutions in Fig.~\ref{Fig:numericalP} are used, with the same color-coding. }
\label{Fig:study}
\end{figure}

We set-up the configuration of the string by assuming that its extremes are
attached to the brane at $\r=\r_1\gg 1$, and treat this as a UV cut-off.
In the numerical study and in the resulting plots, we used $\r_{1}=30$.
The string is stretched in the Minkowski direction $x=x(\r)$, with $x(\r_1)=0$ for convenience.
We vary the integration constant $C^2 > f^2(0)$. For each choice of $C$ 
we  define $\r_0$ as $V^2_{eff}(\r_0)=0$. In this way, the coordinates of the string are
$(x(\r),\r)$,  where
\beqs
x(\r)&=&\int_{\r}^{\r_1} \frac{\di r}{{V_{eff}(r)}}\,.  
\eeqs
The Minkowski distance between the the end-points of the string is hence $L_{QQ}=2x(\r_0)$.
For the energy, because in the numerical study we do not remove the UV cut-off,
rather that Eq.~(\ref{EL}), we use the unsubtracted 
 action (setting $T/(2\pi\alpha^{\prime})=1$)
evaluated up to the cut-off:
\beqs
E_{QQ}&=&2 \int_{\r_0}^{\r_1}\di r \sqrt{\frac{f^2(r)g^2(r)}{f^2(r)-C^2}}\,.
\eeqs

\begin{figure}[htb]
\centerline{\includegraphics[width=10.1cm]{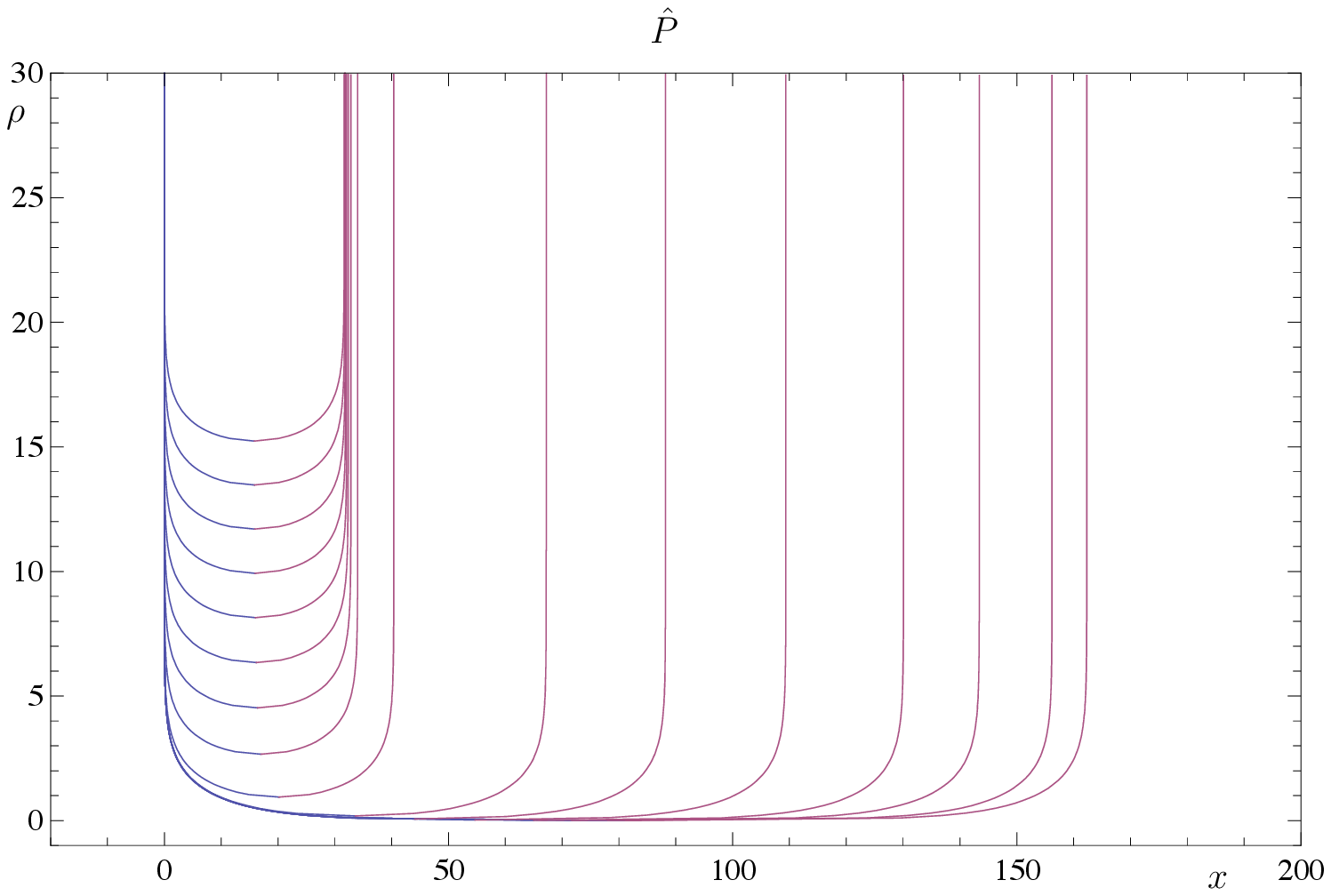}}
\centerline{\includegraphics[width=10.1cm]{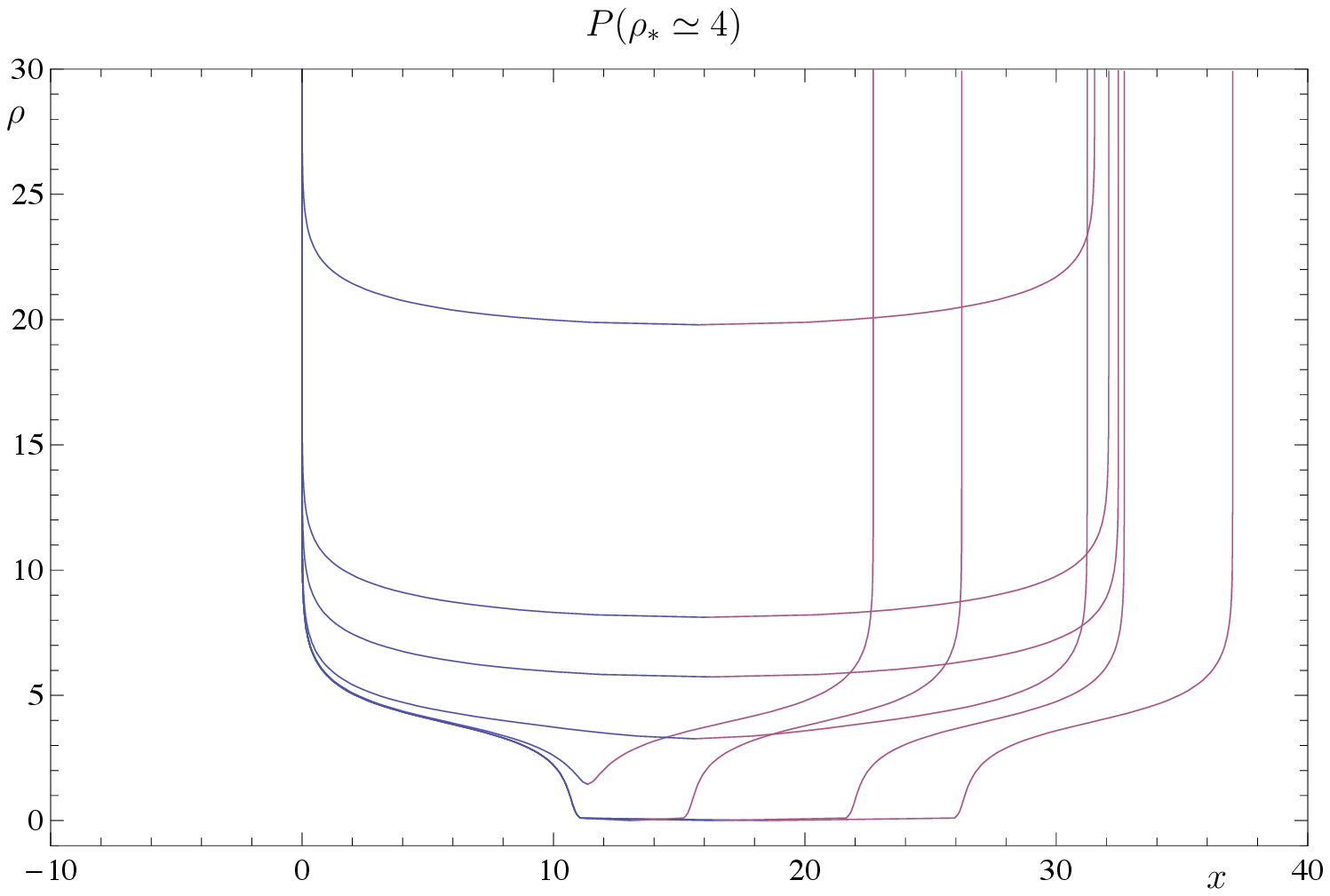}}
\centerline{\includegraphics[width=10.1cm]{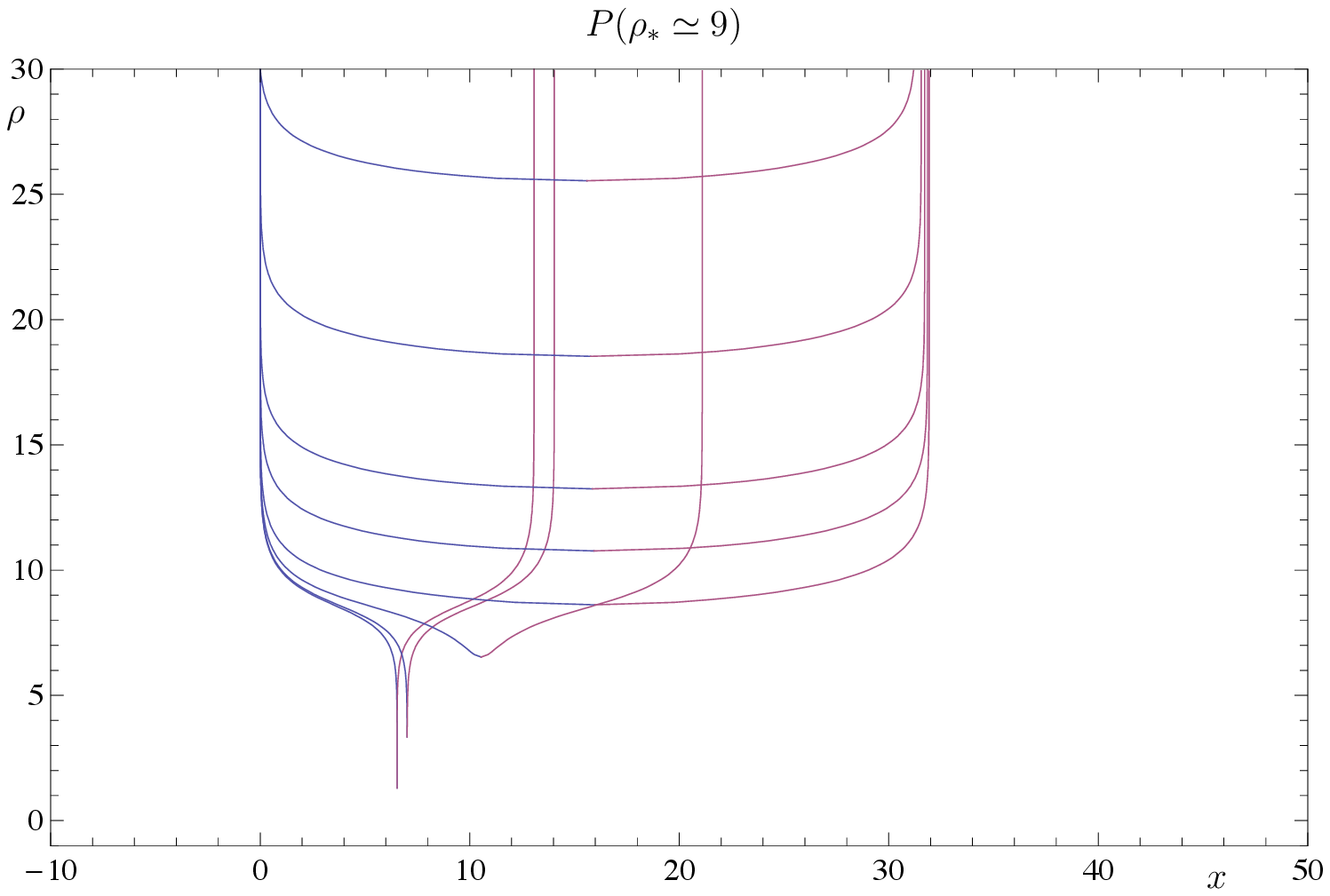}}
\caption{The strings in  $(x,\r)$-plane, 
obtained with various choices of $C^2>f^2(0)$.
 Top to bottom, the three numerical solutions corresponding to 
increasing values of $\r_{\ast}$}
\label{Fig:strings}
\end{figure}

The numerical results obtained for the three different solutions $P$ 
are shown in Fig.~\ref{Fig:study} and Fig.~\ref{Fig:strings}.

Let us first focus our attention on the $\hat{P}$ case of 
Eq.~(\ref{Eq:MN}).
The deeper the string probes the radial coordinate 
(smaller values of $\r_0$), the longer the separation $L_{QQ}$ between the 
end-points on the UV brane, in agreement with our criteria and results of 
sections \ref{general} and \ref{examplesunderstood}.

Two regimes can be identified: as long as $\r_0 > \r_{IR}$, 
then $L_{QQ}$ varies very little with $\r_0$,
while for small $\r_0$, further reductions of $\r_0$ imply much longer $L_{QQ}$.
The scale $\r_{IR}\sim {\cal O}(1)$ is the scale in which the function $Q$ changes from linear
to approximately quadratic in $\r$, and is also the scale below which the gaugino condensate
is appearing (the function $b(\r)$ in the background is non-zero).
This result is better visible in the upper panel of Fig.~\ref{Fig:study}.
The dependence of $L_{QQ}$ on $\r_0$ is monotonic, but shows two very different 
behaviors  for $\r_0<\r_{IR}$ and $\r_0>\r_{IR}$, respectively. The transition between the two
is completely smooth.

The physical meaning of this behavior is well illustrated by studying the total 
energy $E_{QQ}$ of the classical configurations, as a function of $L_{QQ}$ and of $\r_0$
(see the middle and lower panel of Fig.~\ref{Fig:study}).
One sees that for small $L_{QQ}$, the energy 
grows  very fast with $L_{QQ}$, until a critical value beyond which 
the dependence becomes linear. 
We have already studied 
analytically this 
behavior, which can be interpreted in terms of the linear behavior of the
quark-antiquark potential obtained from the Wilson loop in agreement with the discussion of 
sections \ref{general} and \ref{examplesunderstood}.
The energy is also a monotonic function of $\r_0$.

By comparing with the solutions that {\it walk} in the IR,
one sees that a very different behavior appears.
Starting from the upper panel in Fig.~\ref{Fig:study}, one sees that 
as long as $\r_0>\r_{\ast}$, the dependence of $L_{QQ}$ from $\r_0$ reproduces the $\hat{P}$
case. Beginning from such large $\r_0$, we start pulling the string down to smaller values of $\r_0$,
and follow the classical evolution.
 Provided we do this
 adiabatically, we can describe the motion of the string as the set of 
 classical equilibrium solutions we obtained in the previous section.
Going to smaller $\r_0$,  $L_{QQ}$ increses, 
 and nothing special happens
 until the tip of the string touches $\r_0\simeq \r_{\ast}$.
 At this point $L_{QQ}=L_{max}$.
 From here on, the string can keep probing smaller values of $\r_0$ only at the price of becoming shorter in the Minkowsi direction (smaller $L_{QQ}$).
Another change happens when $L_{QQ}=L_{min}$, at which point the 
  tip of the string 
entered the bottom section  of the space,
$\r_0<\r_{IR}$.
 From here on, further reducing $\r_0$
requires larger values of $L_{QQ}$. Asymptotically for 
$\r_0\rightarrow 0$, the separation between the end-points of the string is 
diverging, 
 $L_{QQ}\rightarrow \infty$.

Even more interesting is the behavior of the energy (lower panel of Fig.~\ref{Fig:study}): 
for very short $L_{QQ}$, and again for very large-$L_{QQ}$,
 it is just a monotonic function, but for a range $L_{min}<L_{QQ}<L_{max}$ there are three different configurations allowed
by the classical equations for the string we are studying~\footnote{
Notice that the qualitative behavior of the solutions with $\r_{\ast}\simeq 4$ and $\r_{\ast}\simeq 9$
are identical. However, we were not able to follow numerically the solution with larger $\r_{\ast}$,
and hence the plots show only two such solution. The existence of the third is
assured by the fact that this background must yield a confining potential.}.
One of the three solutions (smoothly connected to the small-$L_{QQ}$ configurations) 
is just the Coulombic potential already seen with $\hat{P}$.
The highest energy one is an unstable configuration, with much higher energy.
The third solution (smoothly connected 
to the unique solution with $L_{QQ}>L_{max}$) reproduces the linear potential typical of confinement.
 Notice (from the lower panel of Fig.~\ref{Fig:study})
that the solution at large-$L_{QQ}$ is linear, but has a slope much larger than what seen in 
the $\hat{P}$ case.
This co-existence of several disjoint classical solutions is expected in 
systems leading to phase transitions (see Appendix~\ref{vanderwaals}).
 The instability/metastability/instability of the 
 solutions can be illustrated by comparing Fig.~\ref{Fig:study} with Fig.~\ref{Fig:vdW} 
 in  Appendix~\ref{vanderwaals}. 
 This is just an analogy, and one should not push it too far. However, 
 identifying the pressure $P$, volume $V$ and Gibbs free energy $G$
as $L_{QQ}\leftrightarrow P$, $\r_0\leftrightarrow V$
 and $E_{QQ} \leftrightarrow G$, one sees obvious similarities.
In particular, there is a critical distance $L_{min}<L_c<L_{max}$
at which the minimum of $E_{QQ}$  is not differentiable.

In order to better understand and characterize the solutions we find,
it is useful to look more in details at the shape of the string configurations,
focusing in particular on the middle panel
in Fig.~\ref{Fig:strings}, in which we plot the string configuration 
that solves the equations of motion for various values of $\r_0$,
on the background with $\r_{\ast}\simeq 4$.
Consider those strings that penetrate below $\r_{\ast}$.
Besides having a shorter $L_{QQ}$, and higher $E_{QQ}$
than those for which $\r_0>\r_{\ast}$, this strings show another interesting feature.
They start developing a non trivial structure around their middle point, 
that becomes progressively more curved
the further the string falls at small $\r$.
Ultimately, this degenerates into a 
cusp-like configuration, which disappears once $\r_0$ approaches the end of the space.
Notice that, as a result, the three different solutions for $L_{min}<L_{QQ}<L_{max}$
have three very different geometric configurations. One (stable or metastable) configuration
is completely featureless, and practically indistinguishable from the solutions in the background
generated by $\hat{P}$.
The second (stable or metastable) configuration shows a funnel-like structure below $\rho_{\ast}$,
and then the string lies very close to the end of the space.
The third (unstable) solution presents a highly curved configuration around its middle point~\footnote{
This should not be interpreted literally as a cusp. 
The classical solutions we found are always continuous and differentiable, and this structure disappears 
once the string approaches the end of the space.}.
All of this seems to be consistent with the fact that in the region $\r_{IR}<\r<\r_{\ast}$
the background has a higher curvature, and as a result the classical configurations prefer 
to lie either in the far-UV or deep-IR. It must be 
noted here that the Ricci curvature is indeed bigger
in this region, but that it converges to a constant for $\r\rightarrow 0$.

\subsection{Comments on this section.}

We conclude this section by summarizing here three 
important  lessons we learned.

First,  we are able to derive the linear $E_{QQ}(L_{QQ})$
expected from confinement. This linear behavior 
emerges for $L_{QQ}\gg L_c$, $L_c$ being the critical distance
between the end-points of the string, 
of the order of the distance $L_{QQ}$ computed
for the string the tip of which reaches $\r_{IR}$.
This fact provides a physical meaning for the 
scale $\r_{IR}$, which clearly shows in the (scheme-dependent) 
background functions and in the gauge coupling \footnote{A change of 
scheme in the string picture corresponds to a redefinition of the 
radial coordinate.}.

In particular, we can conclude that the walking solution we
looked at is dual to a confining theory, but a very different one from the
one of the non-walking solution $\hat{P}$.
This is signaled by the fact that the leading order behavior of the walking solution is related to the value of $V_{eff}$ in $\rho_0\gtrsim\rho_{\ast}$ while the non-walking is related to the zero of $V_{eff}$ in $\rho_0\gtrsim0$, and hence the presence of this two different solutions has to be related to the existence of two different scales in the background.

A limitation of this formalism is that, as we explained at length, 
it does not allow to study the sub-leading corrections to the linear behavior, which
would require a treatment in which $\alpha^{\prime}$-suppressed quantum corrections are included.
It would be very interesting to calculate these corrections, which might provide useful information
as to the nature of the dual theory.

 The second important lesson we learn is related to the second scale $\r_{\ast}$ appearing 
 in solutions with walking behavior. Again, this scale appears in $P$ and as  a
 consequence in all the functions in the background, including the
 (scheme-dependent) gauge coupling.
 The main point is that, provided $\rho_{\ast} > \r_{IR}$,
 $\rho_{\ast}$ has a very important physical meaning: it separates the small-$L_{QQ}$
 regime, where the background and all physical quantities are the same as in the original $\hat{P}$ solution,
 from the large-$L_{QQ}$ regime, in which the dual theory is completely different from the dual to the $\hat{P}$ solution.
 The scale $\rho_{\ast}$ is not just a scheme-dependent fluke effect: its value is somehow related to the coefficient
 of the linear leading behavior of the quark-antiquark potential.

The third lesson we learn is that in the region $\r_{IR}<\r<\r_{\ast}$, the dynamics being
more strongly coupled than elsewhere,  classical solutions are unstable.
The string configurations that are stable are those that either do not reach $\r_{\ast}$,
or those that go through this region only in order to reach the region near the end of the space, where the string can 
lie up to indefinitely large separations (in the Minkowski directions).
This is a very interesting result, that might be related with what found in~\cite{Elander:2009pk},
where it is shown that for large values of $\r_{\ast}$ the spectrum of scalar glueballs 
on these backgrounds splits into a set of  towers of  heavy states and some light state,
separated by a large gap. This is going to be studied elsewhere~\cite{ENP}.

%
\section{Wilson Loop in a Field Theory with 
Flavors}\label{flavoredbackgroundsxx}
In this section we consider the effects of fundamental (flavor) 
degrees of freedom on the Wilson loop, using backgrounds that encode the 
dynamics of fields charged under a flavor group. As discussed in the 
Introduction, 
we expect that confinement does not take place, instead the theory 
will screen. We will observe the existence of a maximal length, 
requiring the string to break. 
As we explained, we are not including $g_s$ effects hence the 
pair-creation will not be accessible to the description we are giving here. 
In~\cite{Karch:2002xe} these effects are taken into account by constructing 
the screened solution explicitly. 

The construction of string 
backgrounds where the effects of flavors is included was considered in a 
large variety of models, see \cite{bunch}. We will concentrate on the 
models developed in 
\cite{Casero:2006pt}.
We follow the treatment and notation of
\cite{HoyosBadajoz:2008fw}, as done in the previous sections. 
Other authors have studied effects similar to the ones 
described in this section by using different string models, see 
\cite{Bigazzi:2008gd}-\cite{Ramallo:2008ew}.

\begin{figure}[htpb]
\begin{center}
\includegraphics[width=12cm]{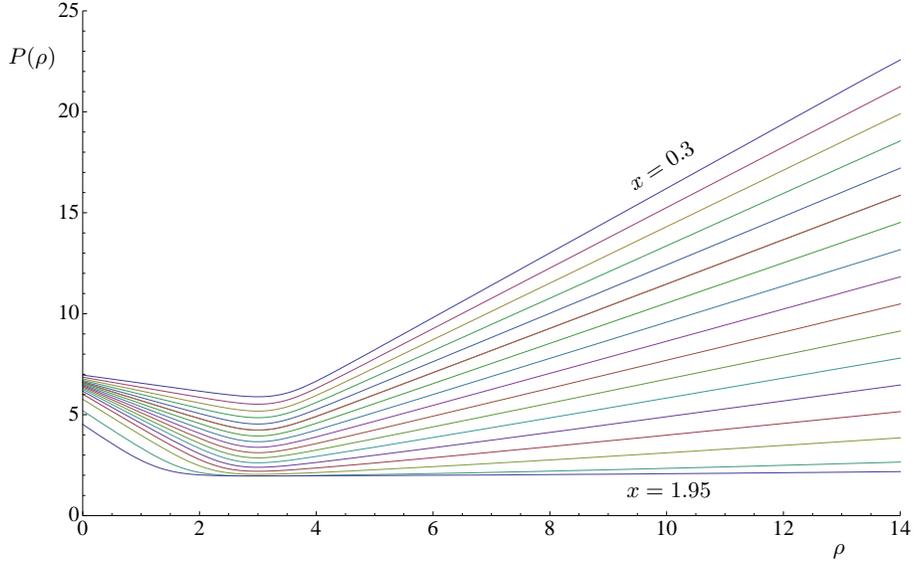}
\end{center}
\caption{Numerical solutions of $P(\r)$ for various values of $N_f/N_c$.}
\label{Fig:PNf}
\end{figure}

At this point, let us stress that, what we will do in the following is to use as probe a closed string
in a classical theory with no $g_s$ correction.
Such an observable cannot describe the physics of the broken $Q\bar{Q}$ pair in the QFT sector; It will have a good overlap with the state describing the $Q\bar{Q}$ pair only for small separation distance among the quarks.
If the distance among the quark is increased another configuration, that will mimic the breaking of the pair, will become more energetically favorable.
On the other hand if we decide to measure only the Wilson loop we will see no signal of such a breaking.
Conversely if a cusps appears in the profile of the string used to measure the wilson loop, this signal has to be read as
the loss of validity of the approximation imposed on our theory.
While this signal cannot be interpreted as corresponding physical quantity (the distance at which the cusp appears has nothing to do with the scale of breaking of the $Q\bar{Q}$ pair), the observation can give a suggestion about the range of validity of our approximation. The results of this section should be understood as the application of the formalism of section~\ref{general} to the case of flavored background, with the proviso of the validity of the Nambu-Goto approximation and the comment made above about the screening phenomena. Aside form this, the numerical solution Fig.~\ref{Fig:PNf} below is new material included here.

In order to find the background solutions, we will need to solve a 
differential equation that is a generalization of the one discussed in the 
previous section, Eq.~(\ref{master}):
\bea
& & P''+ (P'+N_f)\Big[ \frac{P'+Q'+2N_f}{P-Q}+  \frac{P'-Q'+2N_f}{P+Q}   
-4 \coth(2\rho)\Big]=0,\nonumber\\
& & Q(\rho)=\frac{2N_c-N_f}{2}(2\rho \coth(2\rho)-1).
\label{masterflavors}
\eea
The relation to the functions that appear explicitly in the background
is given in Eqs.~(3.18) and (3.19) of  \cite{HoyosBadajoz:2008fw}.

There are various known solutions to  Eq.~(\ref{masterflavors}). We 
focus on the so called type-I solutions that are known only as 
a series expansion. For small values of the radial coordinate ($\rho\to 
0$), the expansion is given in Eq.~(4.24) of \cite{HoyosBadajoz:2008fw},
while for  large values of the radial coordinate ($\rho\to\infty$) is given 
in Eqs.~(4.9)-(4.11)  of \cite{HoyosBadajoz:2008fw}:
\bea
& & P(\rho\sim 0)=P_0- N_f \rho +\frac{4}{3}c^3P_0^2 \rho^3+\cdots\,
,\nonumber\\
& & P(\rho\sim\infty)= Q+ N_c (1+ \frac{N_f}{4Q} +\frac{N_f(N_f-2N_c)}{8 
Q^2}+\cdots)\,.
\label{expansion}
\eea
In the $N_f\to 0$ limit this can be matched with the expansion Eq.~(\ref{Eq:Pexpan}). 
Below, we find it useful to have the IR ($\r\to 0$) asymptotics for the 
functions
\beq
e^{4\phi}\sim \frac{8e^{4\phi_0}}{c^3 P_0^4}\Big[1+\frac{4N_f}{P_0}\r 
+\frac{10 N_f^2}{P_0^2}\r^2   \Big] +....,\;\;\; e^{2k}\sim 2 c^3 
\Big[P_0^2 \r^2 - 2P_0 N_f \r^3  \Big] + ....
\label{irvvv}
\eeq
and in the UV ($\r \to\infty$),
\beq
e^{4\phi}\sim \frac{e^{4\r}}{\r},\;\;\; \;\;\;\; e^{2k}\sim 1 .
\label{uvvvv}
\eeq
We plot in Fig.~\ref{Fig:PNf} a set of numerical solutions, 
which reproduce the asymptotic behaviors in Eq.~(\ref{expansion}),
for several different values of $N_f/N_c$.
The existence of these solutions smoothly joining the asymptotic behaviors has been assumed in \cite{HoyosBadajoz:2008fw}.
Here we are explicitly showing that this assumption is correct.
Numerically, we learn 
that (provided $P_0$ is not too small) the function $P$ is convex, as 
shown in Fig.~\ref{Fig:PNf}. Let us move to the study of string probes in 
these backgrounds.

In the light of the discussion of section \ref{eqsofmotion}, we form the 
combinations
\bea
& & f^2= g_{tt}g_{xx}= 
e^{2\phi(\rho)}=
\frac{\sqrt{8}e^{2\phi_0}\sinh(2\rho)}{\sqrt{(P^2-Q^2)(P'+N_f)}},\;\;\; 
C^2= e^{2\phi(\rho_0)}\nonumber\\
& & g^2= g_{tt}g_{\rho\rho}= 
e^{2\phi(\rho)+2 
k(\rho)}=\frac{\sqrt{2}e^{2\phi_0}\sinh(2\rho)\sqrt{P'+N_f}}{\sqrt{(P^2-Q^2)}},\nonumber\\
& & V_{eff}= \frac{1}{e^{k(\rho)} C}\sqrt{e^{2\phi(\rho)} - 
C^2}= \frac{1}{e^{k(\rho)} }\sqrt{e^{2\phi(\rho) -2\phi(\rho_0)} -1}.
\eea
We will choose for convenience the integration constant $\phi_0$ such that
$8 e^{4\phi_0}= c^3 P_0^4$, hence we will have $\phi(\rho=0)=0$.

Following the discussion  below 
Eq.~(\ref{LVeff}) and using eq.(\ref{uvvvv}), the contribution to $L_{QQ}$ 
of the 
integral for large values of  the radial coordinate goes like
\beq
L_{QQ}(\rho_0\to \infty)\sim \int^{\infty} 
\frac{d\rho}{e^{2\rho}}
\eeq
which is finite. Similarly, using Eq.~(\ref{irvvv}), the lower-end of the 
integral contributes with
\beq
L_{QQ}(\rho_0\to 0)\sim \int_{\rho_0 \to 0}  d\rho \sqrt{\rho},
\eeq
which is itself finite. In this case we can see that in Eq.~(\ref{zzz}) 
the quantity $\gamma=-\frac{1}{2}$, according to the discussion around 
Eq.~(\ref{zzz}) implies a finite $L_{QQ}$. Contrast this with the 
examples
studied in section \ref{examplesunderstood} and \ref{flavoredbackgroundsxx}.

\begin{figure}[ht]
\begin{center}
\includegraphics[width=12cm]{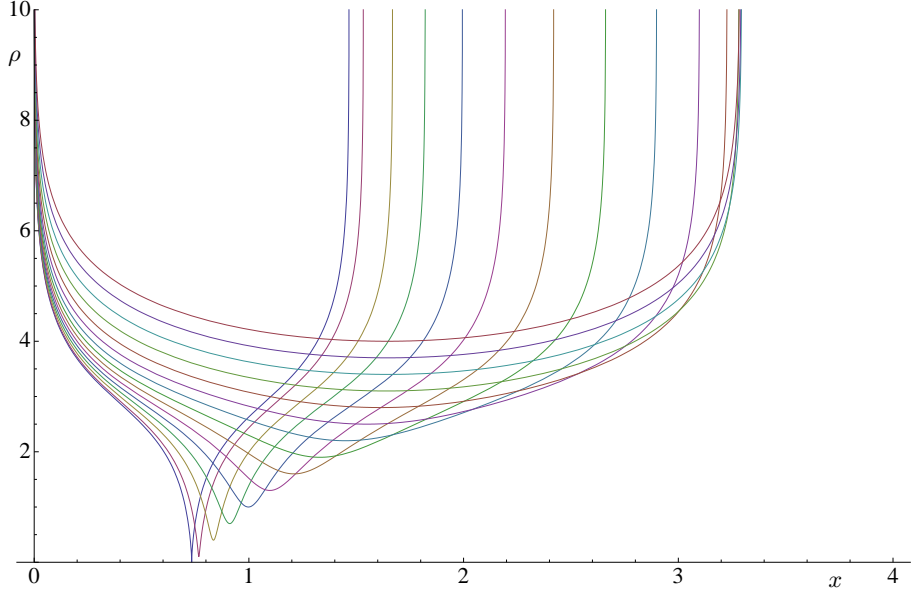}
\caption{Shapes of the string for various choice of $\r_0$}
\label{Fig:shapesNf}
\end{center}
\end{figure}

The fact that near $\rho =0$ the quantity $V_{eff}\sim \r^{-\frac{1}{2}}$ 
implies, using Eq.~(\ref{drdxfinal}), that near the IR  there must be  a 
cusp-like behavior. A plot of the shape of the string makes this clear, see Fig.~\ref{Fig:shapesNf} where  we string probing a background with $N_f/N_c=1.2$.
In contrast with the example studied in section~\ref{walkingsolutionsxx} these backgrounds have a divergent Ricci scalar at $\r=0$, in agreement with the fact that only for $\r_0=0$ the string presents a cusp. This is the only example presented in this paper with a true cusp in the string profile.

\subsection{The case $N_f=2N_c$.}
\begin{figure}[htpb]
\centerline{\includegraphics[width=12cm]{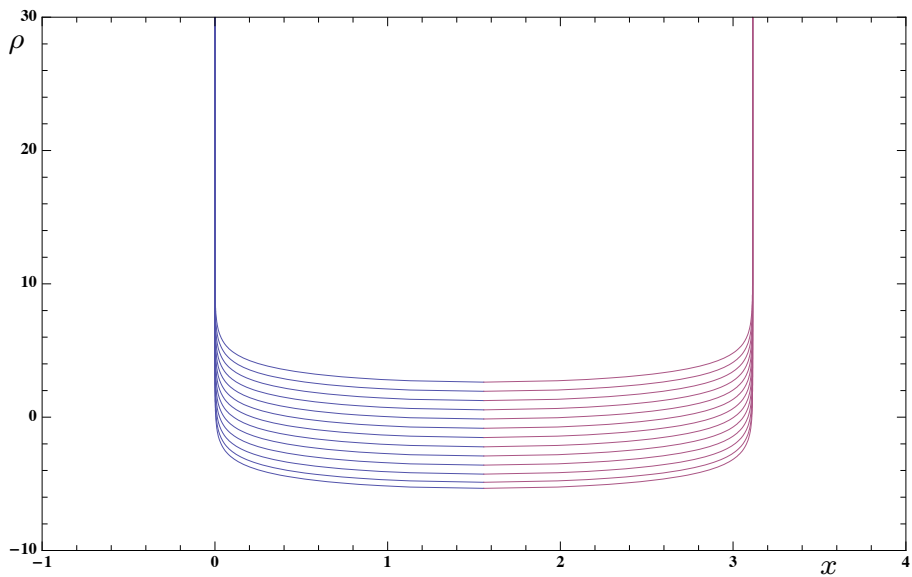}}
\caption{The string hanging on the special background in Eq.~(\ref{Eq:conformal}).
}
\label{Fig:stringconformal}
\end{figure}

Finally, we turn our attention to the very peculiar, limiting  
case with $N_f=2N_c$. The solution is described in Eq.~(4.22) of the 
first paper in \cite{Casero:2006pt}. It corresponds to 
\beqs
Q=4N_c\frac{(2-\xi)}{\xi(4-\xi)}\,, P=N_c +\sqrt{N_c^2+ Q^2}\,,
\label{Eq:conformal}
\eeqs
 and the 
functions $f(\rho), g(\rho)$ read in this case
\beq
f^2=g^2= e^{2\phi_0 +2\rho}
\eeq
where now the radial coordinate is defined in the interval 
$-\infty<\rho<\infty$. The  explicit computation of Eqs.~(\ref{EL}), now 
for  the lower end of the integral $\r_0 \to -\infty$, gives
\beq
L_{QQ}= \sqrt{g_s\alpha' N_c}\pi,\;\;\;\;E_{QQ}=0
\eeq
We illustrate in Fig.~\ref{Fig:stringconformal} the behavior of the string  
on this background.
Notice how the length of the string is fixed 
$L_{QQ}=\pi$, in units where $\alpha^{\prime}g_sN_c=1$,
irrespectively of how deep into
the radial direction the string probes the background. Another interesting 
fact is that the quantity of Eq.~(\ref{rara}) vanishes identically in this 
background, meaning that $dL_{QQ}/d\r_0=0$ for any $\r_0$, implying that it is always possible to find a solution with any given $\r_0$ and with the same $L_{QQ}$, as 
Fig.~\ref{Fig:stringconformal} clearly indicates.
The field theory dual to this background is quite peculiar. On the one hand, the numerology $N_f=2N_c$ brings to mind the situation in which the theory becomes conformal. Nevertheless, we should not forget that this is a background constructed with wrapped branes, so, clearly a scale is introduced (the scale set by the inverse size of the cycle wrapped). It may be the case that once the singularity is resolved one finds in the deep IR an $AdS_5$ space. But the theory should remind that in the UV conformality is broken, hence developing the scale $\Lambda^{-2}=\alpha' g_s N_c$, that is typical of six dimensional theories. Notice that the only allowed separation, according to the calculations above is precisely $\pi$ in units of this scale and the energy is zero. These results do not need to be generic for all solutions with $N_f=2N_c$, but for this particular one discussed above.
Note also that in this case, even when the background is singular in the IR, the string does not become cuspy, as the case studied in the previous section.

\section{Summary and Conclusions}\label{conclusionszz}
Let us summarize what we learnt in this paper and emphasize the reasons 
that motivated our study. In section \ref{general} we recovered many 
well-known results about holographic Wilson loops, but at the same 
time introduced a certain amount of formalism and some new results that allow us to make 
a systematic study of 
many qualitative features of string probes dual to Wilson loops in field 
theory. For example, we introduced  the quantity $V_{eff}$ in 
Eq.~(\ref{nnzz}) which dictates the behavior of the string near the 
boundary of the space ($\r \to \infty$)
and tells us if the Dirichlet boundary condition is attained. At the 
same time, $V_{eff}$ near the IR tells us about the possibility (or not) of 
stretching 
the probe string indefinitely. Assuming certain characteristic power-law  
behavior
of the function $V_{eff}$ near the IR, we were able to derive a condition 
on the power ensuring the presence of a minimum (or turning point).
In that same section, we derived an {\it exact} 
expression of the relation between the energy and the separation of the 
quark pair. We also found  an integral expression giving a necessary 
condition for the existence of turning points in the string probing the 
bulk and the condition for the existence of cusps.

In section \ref{examplesunderstood}, we applied the formalism described 
above to some well-understood examples and made some comments about the 
absence of Luscher terms in the $E_{QQ}-L_{QQ}$  relation.

The material in section~\ref{walkingsolutionsxx} was derived with 
the following motivation:
in the paper 
\cite{Nunez:2008wi} a new background was proposed to be dual to a QFT
with a walking regime (for a particular coupling). Nevertheless, aside 
from an argument based  on 
symmetries 
given in that paper, it could have been the case that this walking region 
is just 
a fluke of the coordinates choice that dissapears under a simple 
diffeomorphism (in the dual field theory, gauge couplings and 
beta functions are scheme dependent). To clarify if there is any real 
physical effect in the walking region we computed Wilson loops in the 
(putative) walking
field theory. In order to do this,  we needed to find a new walking 
solution that allows
the string to satisfy the boundary conditions discussed in section 
\ref{general}. The new solution was found numerically as a perturbation 
of 
the solution in \cite{Maldacena:2000yy}. We gave an interpretation of this 
new background in the dual field theory as the appearance of a VEV for a 
quasi-marginal operator.
When studying the dynamics of strings in this new background we 
observed that there are various physical effects associated with the 
presence and length of the walking regime: the value of the 
Ricci scalar, the separation between the quark pair, 
the relation between the energy and the separation of the pair, etc.
In conclusion, the walking regime has physically observable effects, hence 
it can not erased by a change of coordinates (conversely, it is not an 
effect of a choice of scheme in the field theory). Recent experience 
seems to suggest that whenever we deal with a system with two {\it 
independent} scales, the phenomenology of the Wilson loop will be 
similar to what we described in section \ref{walkingsolutionsxx}.

Finally we moved to the study of the Wilson loops in field theories with 
flavors. Here we benefitted from some work done in the past, where the 
asymptotic behavior of the background functions was given. We constructed 
numerically the full solution in terms of a convenient formalization of the 
problem described in \cite{HoyosBadajoz:2008fw}. We applied the material 
of section \ref{general} to this case and checked that the probe string  
was behaving as expected (screening).

We believe that this paper clarifies  a considerable number of new and 
interesting points. Certainly, it pushes forward the idea of using string 
theory methods to study models of walking dynamics. This may become 
important at the moment of modelling the mechanism  
electroweak symmetry breaking. 
On the phenomenological side we provide a set of tools that can be applied to other backgrounds and may be even used in bottom-up approaches to the dual of QCD.

It would be nice to find new models of walking dynamics and apply the 
ideas in this paper to study and compare features. It would also be 
interesting to apply the formalism developed here to some of the examples 
of backgrounds dual to field theories with flavors, for which certain 
aspects of Wilson loops were studied \cite{Bigazzi:2008gd} - 
\cite{Ramallo:2008ew}.
\newpage
\appendix

\section{UV asymptotic solutions.}
\label{uvasymptotics}
In this appendix we present a short digression about the asymptotic behavior of the solutions to Eq.~(\ref{master}),
in order to make contact with a possible field theory interpretation of the results that have been presented in the section~\ref{walkingsolutionsxx}.
Generically, one can expand all the functions appearing in the background,
and associate the integration constants appearing in this expansion 
with the insertion in the UV of the dual field theory of operators with 
scaling behavior determined by the $\r$-dependence 
of the corresponding term in the expansion.
This is reminiscent of what is done in the case of backgrounds that are
asymptotically AdS, though in the present case the procedure is 
less solid, and the results should be taken with caution.

We can start from the function $Q$. 
\beqs
Q&\sim&N_c\left(2 \r -1 \,+\, {\cal O}(e^{-4\r})\right)\,,
\eeqs
where we dropped factors as powers of $\r$  in the exponential
corrections. We will do so in rest of this analysis, the logic of which will not be affected.

If one thinks that the two behaviors correspond to the insertion of
an operator of dimension $d$ and of its $6-d$ dimensional coupling in the 
(six-dimensional) theory living on the $D5$ branes, this means that 
this expression should be written as
\beqs
Q&\sim&z^d\,+\,z^{6-d}\,,
\eeqs
with $z$ 
proportional to a length scale.
By comparison with the expansion of $Q$, one is lead to
the identification $\r=-\frac{3}{2}\log z$.
This result agrees with what was done by comparing an appropriately defined 
beta-function to the NSVZ beta function in~\cite{DiVecchia:2002ks}.
One can interpret the two terms in this expansion as the deformation of
the theory  by the insertion of a marginally relevant
 operator of dimension $d\sim 6-\epsilon$,
with coupling of dimension $6-d\sim \epsilon$.
Notice that the coefficient of the ${\cal O}(e^{-4\r})$ correction is very small,
and has a sizable effect only at very small values of $\r$. This 
is {\it not} due to the fact that we 
chose a particular value of the integration constant $Q_0$, in order to avoid the arising of 
a pathology in the IR. Allowing for $Q_0$ would yield the same expansion, and the
modification of the coefficients would be such that only for $\r<\r_{IR}$ 
would the sub-leading correction have an effect. The scale $\r_{IR} \sim {\cal O}(1)$
has an important role in the present study.

Another important quantity in the background is 
\beqs
b(\r)&=&\frac{2\r}{\sinh(2\r)}\,\sim\,{\cal O}(e^{-2\r})\,,
\eeqs
which by means of the previous identification is equivalent to $b\sim z^3$.
This can be interpreted as the VEV of a dimension-3 operator (customarily identified with
the gaugino condensate in the dimensionally-reduced 4-dimensional low-energy theory).  
This is a relevant deformation.
Because this is a more relevant operator than the VEV mentioned above, 
it always dominates in the IR over the sub-leading correction to $Q$, which can be ignored.
Notice also that  $b(\r)$ is practically vanishing for $\r>\r_{IR}$.
This provides an intuitive explanation for this scale: it is the scale at which the gaugino condensate
emerges dynamically. It also suggests that the expansion of $Q$, rather than identifying 
an independent operator, might be interpreted in terms of the insertion
of an operator loosely corresponding to the square of the gaugino condensate.

There exist two different classes of 
UV-asymptotic solutions for $P$~\cite{HoyosBadajoz:2008fw}:
\beqs
\label{Eq:classI}
P&\sim&2N_c \r \,+\,{\cal O}(e^{-4\r})\,\,\,\,{\rm (class \,I)}\,,\\
P&\sim&{\cal O}(e^{4/3\r})\,+\,{\cal O}(e^{-4/3\r})\,+\,{\cal O}(e^{-8/3\r})\,\,\,\,{\rm (class\, II)}\,.
\label{jenzz}
\eeqs
In class~I, there is only one integration constant, in the ${\cal O}(e^{-4\r})$ term.
Notice that $P$ has the same leading and sub-leading components as $Q$. 
However, the sub-leading correction depends on a free parameter: the corresponding VEV
can be enhanced in such a way that there be a range of $\r$ over which the dynamics is dominated
by this deformation, over the deformation present in $b(\r)$, while the latter 
will become 
important only at very small values of $\r$.

Solutions in the class~II are rather different.
The independent coefficients appear in the ${\cal O}(e^{4/3\r})$ 
and ${\cal O}(e^{-8/3\r})$ terms, and the former cannot be dialed to zero 
independently of the latter (see~\cite{HoyosBadajoz:2008fw} for details).
Using the same identification between $\r$ and $z$,
the leading order component of $P$ scales as $z^{-2}$, and can be interpreted as the insertion of 
a dimension-8 operator in the six-dimensional theory. It is somewhat natural to think that it is related to a
gauge coupling is six-dimensions. 
The presence of sub-leading corrections
that scale as $z^2$, and  $z^4$ suggest that the gravity field  $P$  
should not be interpreted as a simple
operator in the underlying dual dynamics, and that some caution should be 
used.
One might want to interpret the $z^4$ as the insertion of the VEV of a 
four dimensional operator. But it could as well arise from the combination of the 
coupling scaling as $z^{-2}$  and the same VEV of marginal operator scaling as $z^6$ 
that is present in class I.
All of this should not be taken too literally, but provides some guidance in what we do
in the body of the paper, when constructing the class of walking solutions we are interested in. 

\section{Van der Waals gas.\label{vanderwaals}}
Here we summarize some aspect of first order phase transitions that
plays an important conceptual role in the body of the paper. 
By way of example, we remind the reader about the 
classical treatment of the van 
der Waals gas,  in terms of its pressure $P$, temperature $T$ and volume 
$V$ of $N$ moles of particles, by means of the  equation of state 
\beq
P= \frac{NRT}{V- b N} - \frac{N^2 a}{V^2}\,,
\label{vdw}
\eeq
where $R, b, a$ are constants.

\begin{figure}[htpb]
\centerline{\includegraphics[width=7cm]{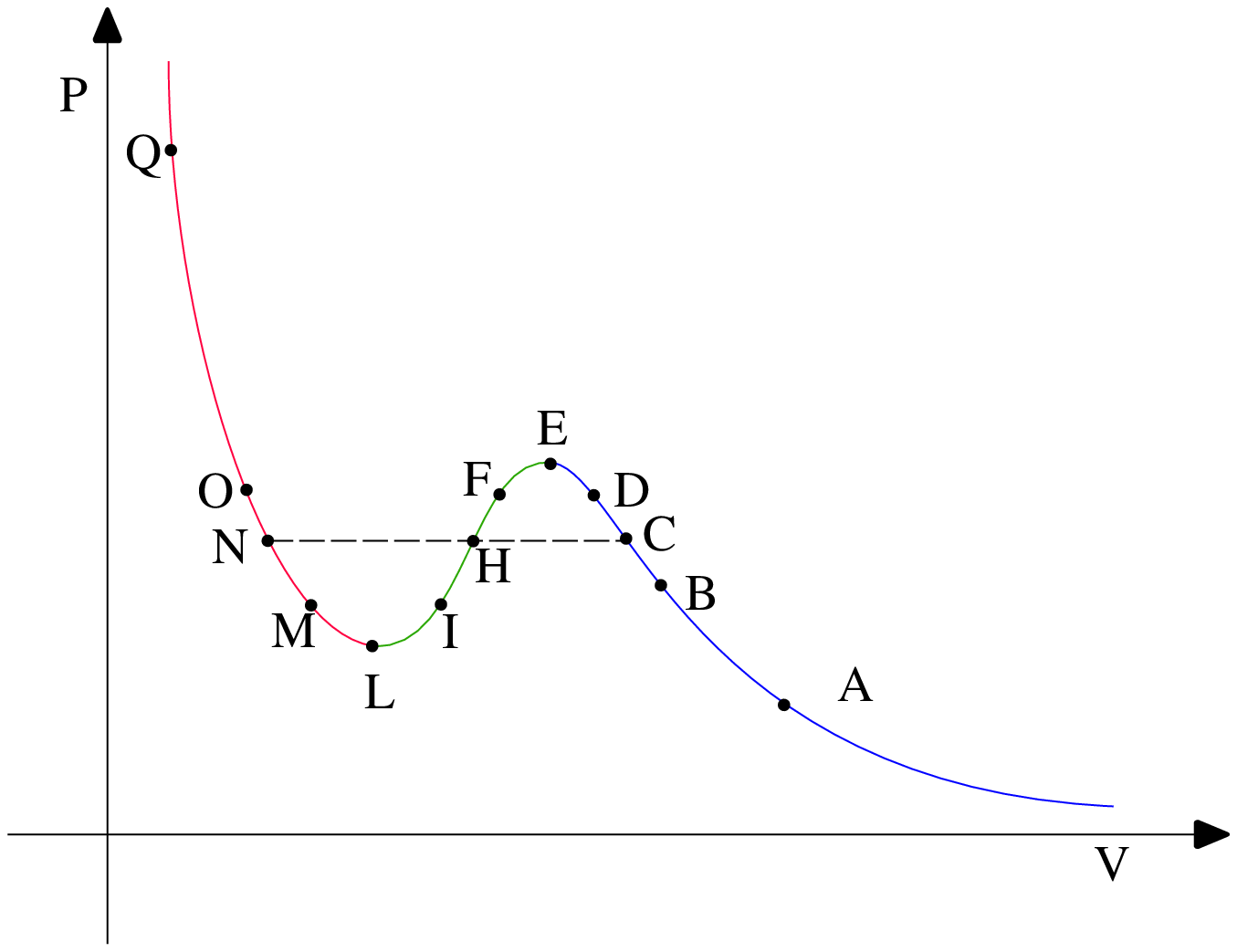}
\includegraphics[width=7cm]{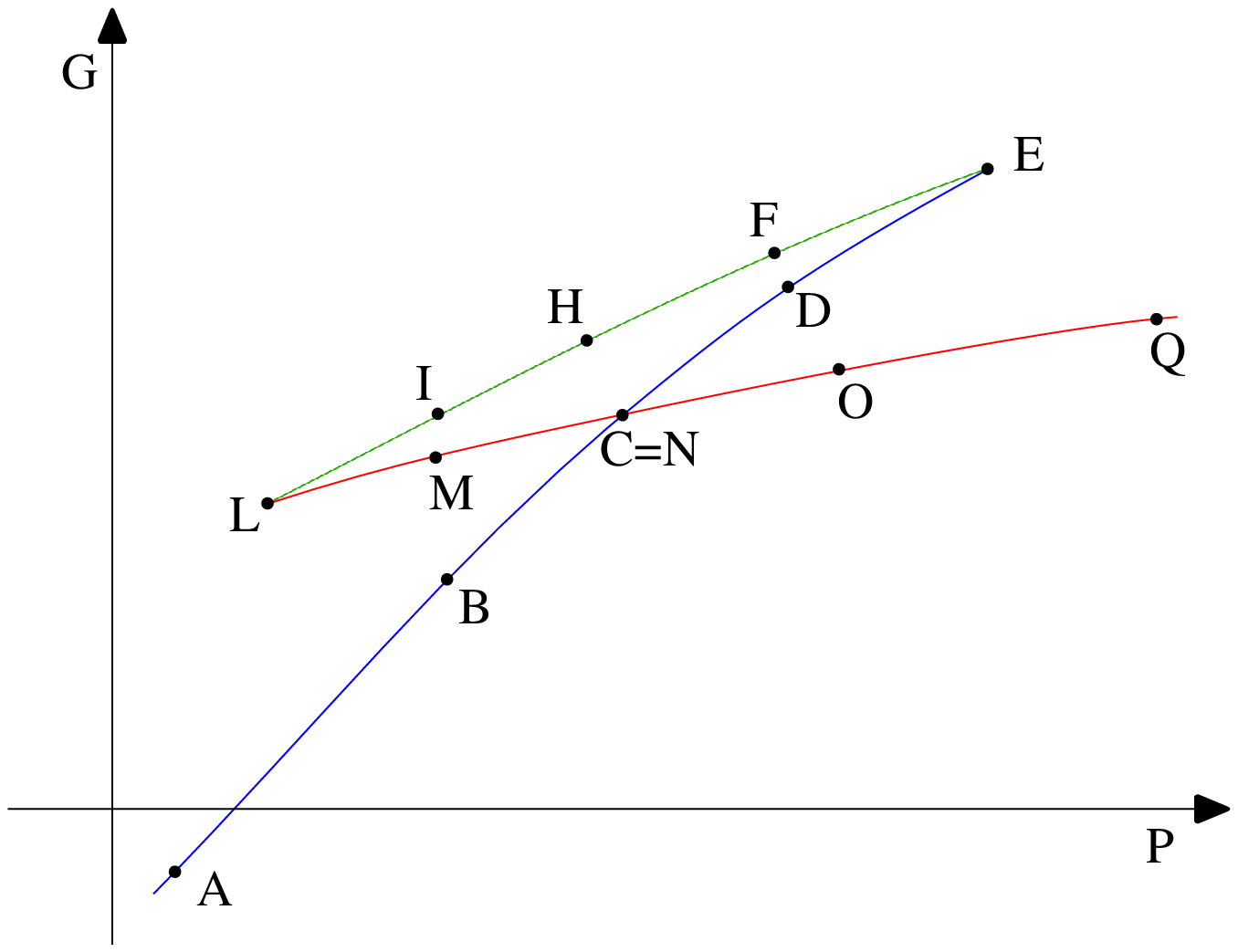}}
\caption{The pressure $P$ as a function of the volume $V$ (left panel)
and the Gibbs free energy $G$ as a function of the pressure $P$(right panel) for
the same isotherm curve.}
\label{Fig:vdW}
\end{figure}

In Fig.~\ref{Fig:vdW}, we plot one isotherm. 
The condition 
for stability of the equilibrium $\left(\partial^2 F /\partial V^2\right)_T = - \left(\frac{\partial P}{\partial V}\right)_{T} >0$ 
is not satisfied in some region. This implies that a phase transition is taking place.

In order to understand what the physical trajectory followed by the
 system at equilibrium is, we plot in Fig.~\ref{Fig:vdW} the  
 Gibbs free energy $G=G(T,P)=G(P)$ for the same isotherm. 
 From this plot, one sees that  the system evolves on the 
path $ABCOQ$, where $C=N$, in such a way that for every choice of $P$
the Gibbs free energy $G$ is at its minimum.
 The evolution is smooth along $ABC$ (gas phase: $|\partial P/\partial V|$ is small), 
 but at $C=N$ the free energy is not
 differentiable, signaling that a first-order phase transition is taking place.
 In the $(P,V)$ plane the system runs along the horizontal line 
 (constant $P$) joining $C$ and $N$.
 This explains  the {\it Maxwell rule} introducing a curve of constant 
pressure that separates two regions of  equal areas above and below it, delimited by the original isotherm.
Afterwards, 
the evolution follows smoothly the curve $NOQ$ (liquid phase: $|\partial P/\partial V|$ is large). 

While the trajectory $ABCNOQ$ follows the stable equilibrium configurations,
it is possible to 
have the system evolving along the path $CDE$ or $LMN$. Both these 
paths represent metastable configurations, because $\left(\partial P/\partial V\right)_T<0$. 
Indeed, these metastable 
states can be realized in laboratory experiments. For example, in a  bubble chamber or in supercooled water,
the metastability is exploited as a detector device, because 
small perturbations induced by passing-by charged particles  are sufficient to
drive the system out of the state and into the stable minimum.
The evolution along the path $EFHIL$ is completely unstable and not realized 
physically (it is a local maximum, as clear from the right panel of Fig.~\ref{Fig:vdW},
and by the fact that $\left(\partial P/\partial V\right)_T>0$).

The analogy  with the examples in the main body 
of the paper is apparent.
Notice for instance that the pressure $P$ in this system as a function of the 
volume $V$ behaves as a non monotonic function, hence there are {\it inversion 
points} in the curve (the points $E,L$ in figure \ref{Fig:vdW}),
in the same sense in which we discuss the presence of inversion points
in the body of the paper.

\newpage
\vspace{1.0cm}
\section*{Acknowledgments}
We wish to thank A. ~Cotrone, F. ~Bigazzi, D.~Elander, S.~P.~Kumar, J. 
Schmude, A.~Paredes for discussions and correspondence.
The work of MP is supported in part  by the Wales Institute of
Mathematical and Computational Sciences.

\end{document}